\documentclass[superscriptaddress,nofootinbib,twocolumn,aps]{revtex4-2}
\usepackage{ulem}
\usepackage{epsfig,amsmath,color,algorithm,algcompatible,amsthm,bm}
\usepackage{multirow}
\usepackage[table,xcdraw]{xcolor}
\usepackage[colorinlistoftodos,prependcaption]{todonotes}

\newcommand{\bold}[1]{{\bf #1}}

\newcommand{\avector}[0]{\bold{a}}
\newcommand{\dvector}[0]{\bold{d}}

\newcommand{\cvector}[0]{\bold{c}}
\newcommand{\cprime}[0]{\bold{c^{\prime}}}
\newcommand{\dprime}[0]{\bold{d^{\prime}}}

\newcommand{\kvector}[0]{\bold{k}}
\newcommand{\data}[0]{|Data\rangle}
\newcommand{\dataket}[0]{\langle Data|}
\newcommand{\stockdim}[0]{{\rm stock}}
\newcommand{\timedim}[0]{{\rm time}}

\newcommand{\E}[2]{\underset{#1}{\bold{E}}[#2]}
\newcommand{\fig}[4]{
\begin{figure*}[ht]
\centering
\includegraphics[#4]{#1}
\caption{#2}
\label{#3}
\end{figure*}
}
\newcommand{\fighalf}[4]{
\begin{figure}
\centering
\includegraphics[#4]{#1}
\caption{#2} 
\label{#3}
\end{figure}
}
\newcommand{\matrixsplit}[1]{\left(\begin{array}{ll}
    {#1}_{\uparrow\uparrow} & {#1}_{\uparrow\downarrow}
    \cr
    {#1}_{\downarrow\uparrow} & {#1}_{\downarrow\downarrow}
    \end{array}
    \right)}
\newcommand{\vectorsplit}[1]{\left(\begin{array}{l}
    {#1}_{\uparrow}
    \cr
    {#1}_{\downarrow}
    \end{array}
    \right)}
\newtheorem{th.}{Theorem}
\newtheorem{co.}{Corollary}
\newtheorem{definition}{Definition}
\newtheorem{innercustomthm}{Theorem}
\newenvironment{customthm}[1]
  {\renewcommand\theinnercustomthm{#1}\innercustomthm}
  {\endinnercustomthm}
\begin{document}

\title{
	Approximate amplitude encoding in shallow parameterized quantum circuits
	\\
	and its application to financial market indicator
}

\author{Kouhei Nakaji}
\affiliation{Graduate School of Science and Technology, Keio University, 3-14-1 Hiyoshi, Kohoku-ku, Yokohama, Kanagawa, 223- 8522, Japan}
\affiliation{Quantum Computing Center, Keio University, 3-14-1 Hiyoshi, Kohoku-ku, Yokohama, Kanagawa, 223-8522, Japan}

\author{Shumpei~Uno}
\affiliation{Mizuho Research \& Technologies, Ltd., 2-3 Kanda-Nishikicho, Chiyoda-ku, Tokyo, 101-8443, Japan}
\affiliation{Quantum Computing Center, Keio University, 3-14-1 Hiyoshi, Kohoku-ku, Yokohama, Kanagawa, 223-8522, Japan}

\author{Yohichi~Suzuki}
\affiliation{Quantum Computing Center, Keio University, 3-14-1 Hiyoshi, Kohoku-ku, Yokohama, Kanagawa, 223-8522, Japan}

\author{Rudy~Raymond}
\affiliation{IBM Quantum, IBM Research-Tokyo, 19-21 Nihonbashi Hakozaki-cho, Chuo-ku, Tokyo, 103-8510, Japan}
\affiliation{Quantum Computing Center, Keio University, 3-14-1 Hiyoshi, Kohoku-ku, Yokohama, Kanagawa, 223-8522, Japan}

\author{Tamiya~Onodera}
\affiliation{IBM Quantum, IBM Research-Tokyo, 19-21 Nihonbashi Hakozaki-cho, Chuo-ku, Tokyo, 103-8510, Japan}
\affiliation{Quantum Computing Center, Keio University, 3-14-1 Hiyoshi, Kohoku-ku, Yokohama, Kanagawa, 223-8522, Japan}

\author{Tomoki~Tanaka}
\affiliation{Mitsubishi UFJ Financial Group, Inc.~and~MUFG~Bank,~Ltd.,\\ 2-7-1 Marunouchi, Chiyoda-ku, Tokyo, 100-8388, Japan}
\affiliation{Quantum Computing Center, Keio University, 3-14-1 Hiyoshi, Kohoku-ku, Yokohama, Kanagawa, 223-8522, Japan}
\affiliation{Graduate School of Science and Technology, Keio University, 3-14-1 Hiyoshi, Kohoku-ku, Yokohama, Kanagawa, 223- 8522, Japan}

\author{Hiroyuki~Tezuka}
\affiliation{Sony Group Corporation, 1-7-1 Konan, Minato-ku, Tokyo, 108-0075, Japan}
\affiliation{Quantum Computing Center, Keio University, 3-14-1 Hiyoshi, Kohoku-ku, Yokohama, Kanagawa, 223-8522, Japan}
\affiliation{Graduate School of Science and Technology, Keio University, 3-14-1 Hiyoshi, Kohoku-ku, Yokohama, Kanagawa, 223- 8522, Japan}

\author{Naoki~Mitsuda}
\affiliation{Sumitomo Mitsui Trust Bank, Ltd., 1-4-1, Marunouchi, Chiyoda-ku, Tokyo, 100-8233, Japan}
\affiliation{Quantum Computing Center, Keio University, 3-14-1 Hiyoshi, Kohoku-ku, Yokohama, Kanagawa, 223-8522, Japan}

\author{Naoki~Yamamoto\thanks{
		e-mail address: \texttt{yamamoto@appi.keio.ac.jp}
	}}
\affiliation{Quantum Computing Center, Keio University, 3-14-1 Hiyoshi, Kohoku-ku, Yokohama, Kanagawa, 223-8522, Japan}
\affiliation{Department of Applied Physics and Physico-Informatics, Keio University, Hiyoshi 3-14-1, Kohoku-ku, Yokohama 223-8522, Japan}

\begin{abstract}
	Efficient methods for loading given classical data into quantum circuits are essential
	for various quantum algorithms.
	In this paper, we propose an algorithm called {\it Approximate Amplitude Encoding}
	that can effectively load all the components of a given real-valued data vector into
	the amplitude of quantum state, while the previous proposal can only load the absolute
	values of those components.
	The key of our algorithm is to variationally train a shallow parameterized quantum
	circuit, using the results of two types of measurement; the standard computational-basis measurement plus the measurement in the Hadamard-transformed basis, introduced in order
	to handle the sign of the data components.
	The variational algorithm changes the circuit parameters so as to minimize the sum of
	two costs corresponding to those two measurement basis, both of which are given by the
	efficiently-computable maximum mean discrepancy.
	We also consider the problem of constructing the singular value decomposition entropy
	via the stock market dataset to give a financial market indicator; a quantum algorithm (the variational singular value decomposition algorithm) is known to produce a solution faster than classical, which yet requires the sign-dependent amplitude encoding.
	We demonstrate, with an in-depth numerical analysis, that our algorithm realizes
	loading of time-series of real stock prices on quantum state with small approximation
	error, and thereby it enables constructing an indicator of the financial market based on the stock prices.
\end{abstract}


\maketitle

\section{Introduction}
\label{SECTION-introduction}

Quantum computing is expected to solve problems that cannot be solved efficiently
by any classical means.
The promising quantum algorithms are Shor's factoring algorithm
\cite{Shor1994AlgorithmsFQ} and Grover search algorithm \cite{Grover1996AFQ}. 
After these landmark findings, a number of quantum algorithms have been developed, 
including quantum-enhanced linear algebra solvers 
\cite{harrow2009quantum,jacob16,siddarth18,maria16,vojtech18,carlo17,park20,maria18,
rebentrost2014quantum,kerenidis2016quantum,wiebe2012quantum,
schuld2017implementing}.
An important caveat is that those algorithms assume that the classical data 
(i.e., elements of the linear equation) has been loaded into the (real) amplitude 
of a quantum state, i.e., {\it amplitude encoding}. 
However, to realize the amplitude encoding without ancillary qubits, in general 
we are required to operate quantum circuit with exponential depth with respect to 
the number of qubits \cite{plesch2011quantum,shende2006synthesis,grover2002creating,Mttnen2005TransformationOQ,Shende2005QuantumCF,A1,A3,A4,A6,Bergholm_2005,A10,B1}. 
Hence there have been developed several techniques to achieve amplitude encoding 
without using an exponential depth circuit, e.g., the method introducing 
ancillary qubits \cite{D1,B2,B3,B4}, which however may introduce an exponential 
number of the ancillary qubits in the worst case. 
References~\cite{E1,E2,E3,E4} achieve this purpose by limiting the data to a 
unary one. 
Also, Refs.~\cite{grover2000synthesis,sanders2019black,C4,C5,C6} employ the 
black-box oracle approach. 
Note that these are perfect or precision-guaranteed data loading methods, 
which consequently can require a large quantum circuit involving hard-to-implement 
gate operations. 
On the other hand, there are many problems that only need an approximate 
calculation (e.g., calculation of a financial market indicator mentioned below), 
in which case it is reasonable to seek an approximate data loading method that 
effectively runs even on currently-available shallow and limited-structural 
quantum circuit.

In this paper, we propose an algorithm called the {\it approximate amplitude 
encoding (AAE)} that trains a shallow parameterized quantum circuit (PQC) to 
approximate the ideal exact data loading process. 
Note that, because of unavoidable approximation error, the application of AAE 
must be the one that allows imperfection in the focused quantity, such as 
a global trend of the financial market indicator which will be described later. 
That is, the scope of this paper is not to propose a perfect or 
precision-guaranteed encoding algorithm. 
Rather, AAE is a data loading algorithm that works with fewer gates, despite 
the unavoidable error caused by the limited representation ability of a fixed 
ansatz and possibly the incomplete optimization.

To describe the problem more precisely, let $\data$ be the target $n$-qubit state whose amplitude represents the classical data component.
Then AAE provides the training policy of a PQC represented by the unitary $U(\boldsymbol{\theta})$, so that the finally obtained $U(\boldsymbol{\theta})$ followed by
another shallow circuit $V$ approximately generates $\data$, with the help of an
auxiliary qubit.
Namely, as a result of training, $VU(\boldsymbol{\theta})|0\rangle^{\otimes n}|0\rangle$  approximates the state $e^{i\alpha}\data|y\rangle$, where
$|y\rangle$ is a state of the auxiliary qubit and $e^{i\alpha}$ is the global 
phase. 
Hence, though there appears an approximation error, the $O(1)\sim O(\mbox{poly}(n))$-depth quantum
circuit $VU(\boldsymbol{\theta})$ achieves the approximate data loading, instead of the ideal
exponential-depth circuit \cite{plesch2011quantum,shende2006synthesis,grover2002creating,Mttnen2005TransformationOQ,Shende2005QuantumCF,A1,A3,A4,A6,Bergholm_2005,A10,B1}.
It should also be mentioned that the number of classical bits for storing $2^n$ 
dimensional vector data is $O(2^n)$, while the number of qubits for this purpose 
is $O(\mbox{poly}(n))$ in several data loading algorithms \cite{plesch2011quantum,shende2006synthesis,grover2002creating,Mttnen2005TransformationOQ,Shende2005QuantumCF,A1,A3,A4,A6,Bergholm_2005,A10,B1,grover2000synthesis,sanders2019black,C4,C5,C6,E1,E2,E3,E4} and the proposed AAE is included in this class.

Our work is motivated by Ref.~\cite{zoufal2019quantum} that proposed a variational
algorithm for constructing an approximate quantum data-receiver, in the framework
of generative adversarial network (GAN);
the idea is to train a shallow PQC so that the absolute values of the amplitude of
the final state approximate the absolute values of the data vector components.
Hence the method is limited to the case where the sign of the data components
does not matter or the case where the data is given by a probability vector as
in the setting of \cite{zoufal2019quantum}.
In other words, this method cannot be applied to a quantum algorithm based on
the amplitude encoding that needs loading the classical data onto the quantum 
state without dropping their sign.
In contrast, our proposed method can encode the sign in addition to the 
absolute value, although there may be an approximation error between the 
generated state and the target state.

As another contribution of this paper, we show that the combination of our AAE algorithm and the
variational {\it quantum Singular Value Decomposition (qSVD)} algorithm \cite{bravo2020quantum}
offers a new quantum algorithm for computing the {\it SVD entropy} for stock
price dynamics \cite{caraiani2014predictive}, which is used as a good indicator 
of the financial market. 
In fact, this algorithm requires that the signs of the stock price data is 
correctly loaded into the quantum amplitudes; on the other hand, the goal is 
to capture a global trend of the SVD entropy over time rather than its precise 
values, meaning that this problem satisfies the basic requirement of AAE 
described in the second paragraph. 
We give an in-depth numerical simulation with a set of real stock price data, to demonstrate that this
algorithm generates a good approximating solution of the correct SVD entropy.

The rest of the paper is organized as follows. 
In Section~\ref{SECTION-algorithm}, we describe the algorithm of AAE.  Section~\ref{SECTION-experiment} gives a demonstration of AAE applied to 
approximately compute the SVD entropy for stock market dynamics. 
Finally, we conclude the paper with some remarks in Section~\ref{SECTION-conclusion}.

\section{Approximate Amplitude Encoding Algorithm}
\label{SECTION-algorithm}

\subsection{The goal of the AAE algorithm}
\label{SECTION-goal}

In quantum algorithms that process a classical data represented by a real-valued
$N$-dimensional vector $\dvector$, first it has to be encoded into the quantum state;
a particular encoding that can potentially be linked to quantum advantage is to encode
$\dvector$ to the amplitude of an $n$-qubits state $\data$.
More specifically, given $|j\rangle$ as $|j\rangle = |j_1 j_2 \cdots j_n\rangle$
where $j_k$ is the state of the $k$-th qubit in computational basis and
$j = \sum_{k=1}^n 2^{n-k} j_k$, the data quantum state is given by
\begin{equation}
	\label{EQUATION-target-state}
	\data = \sum_{j=0}^{N-1}\dvector_j |j\rangle,
\end{equation}
where $N=2^n$ and $\dvector_j$ denotes the $j$-th element of the vector $\dvector$.
Also here $\dvector$ is normalized; $\sum_j \dvector_j^2=1$.
Recall that, even when all the elements of $\dvector$ are fully accessible, in
general, a quantum circuit for generating the state (\ref{EQUATION-target-state})
requires an exponential number of gates, which might destroy the quantum advantage \cite{plesch2011quantum,shende2006synthesis,grover2002creating,Mttnen2005TransformationOQ,Shende2005QuantumCF,A1,A3,A4,A6,Bergholm_2005,A10,B1}.

In contrast, our algorithm uses a $\ell$-depth PQC (hence composed of $O(\ell n)$
gates) to try to approximate the ideal state \eqref{EQUATION-target-state}. The depth $\ell$ is set to be $O(1)\sim O(\mbox{poly}(n))$.
Suppose now that, given an $N$-dimensional vector $\avector$, the state generated
by a PQC, represented by the unitary matrix $U(\boldsymbol{\theta})$ with $\boldsymbol{\theta}$ the
vector of parameters, is given by $U(\boldsymbol{\theta})|0\rangle^{\otimes n} =
	\sum_{j=0}^{N-1} \avector_j|j\rangle$.
If the probability to have $|j\rangle$ as a result of measurement in the computational
basis is $\dvector^2_j$, this means $|\avector_j| = |\dvector_j|$ for all $j$.
Therefore, if only the absolute values of the amplitudes are necessary in a quantum
algorithm after the data loading as in the case of \cite{zoufal2019quantum}, the
goal is to train $U(\boldsymbol{\theta})$ so that the following condition is satisfied;
\begin{equation}
	\label{EQUATION-naive}
	|\avector_j|^2 = |\langle j|U(\boldsymbol{\theta})|0\rangle|^2 = \dvector_j^2, ~
	\forall j\in [0,1,\cdots ,N-1].
\end{equation}
However, some quantum algorithms need a quantum state containing $\dvector_j$ 
itself, rather than $\dvector_j^2$.
Naively, hence, the goal is to train $U(\boldsymbol{\theta})$ so that
$U(\boldsymbol{\theta})|0\rangle^{\otimes n} = \data$.
But as will be discussed later, in general we need an auxiliary qubit and
thereby aim to train $U(\boldsymbol{\theta})$ so that
\begin{equation}
	\label{EQUATION-goal}
	V U(\boldsymbol{\theta})|0\rangle^{\otimes n}|0\rangle
	= e^{i\alpha}\data |y\rangle,
\end{equation}
where $V$ represents a fixed operator containing post-selection and $e^{i\alpha}$
is the global phase.
$|0\rangle$ in the left hand side and $|y\rangle$ in the right hand are the auxiliary qubit state, which might
not be necessary in a particular case (Case 1 shown later).
This is the goal of the proposed AAE algorithm.
When Eq.~(\ref{EQUATION-goal}) is satisfied, the first $n$-qubits of
$V U(\boldsymbol{\theta})|0\rangle^{\otimes n}$ serve as an input of
the subsequent quantum algorithm.

In the following, we assume that all matrix components of $U(\boldsymbol{\theta})$ are real in the computational basis
for any $\boldsymbol{\theta}$, to ensure that $U(\boldsymbol{\theta})|0\rangle^{\otimes n}$ only generates real amplitude quantum states.
In particular, we take $U(\boldsymbol{\theta})$ composed of only the parameterized $R_y$ rotational gate
$R_y(\theta_r)=\exp(-i \theta_r \sigma_y /2)$ and CNOT gates;
here $\theta_r$ is the $r$-th element of $\boldsymbol{\theta}$ and $\sigma_y$ is the Pauli $Y$ operator.
There is still a huge freedom for constructing the PQC as a sequence of $R_y$ and CNOT gates,
but 
in this paper we take the so-called hardware efficient ansatz \cite{kandala2017hardware} due to its high expressibility, or rich state generation capability.
We show the example of the structure of the hardware efficient ansatz in Fig.~\ref{FIGURE-data-circuit}. 
Note that according to the literature \cite{nakaji2021expressibility}, the alternating layered ansatz proposed in \cite{cerezo2021cost} also has high expressibility comparable to the hardware efficient ansatz; thus, the alternating layered ansatz is another viable ansatz for our problem.

\fig{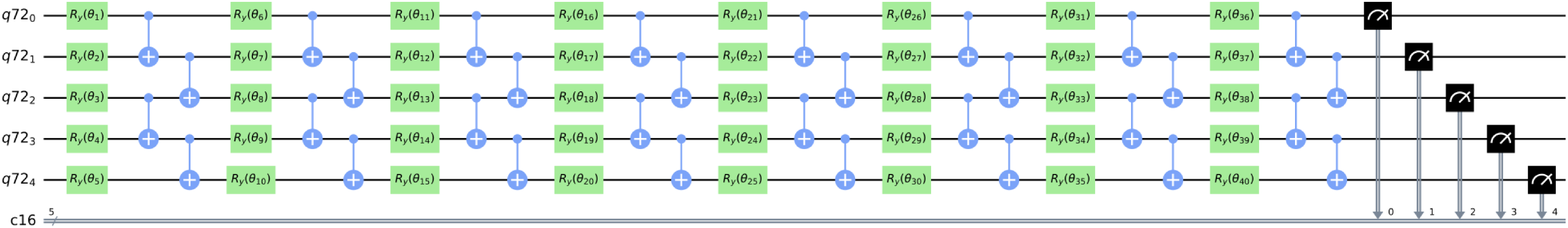}{
Example of the structure of the hardware efficient ansatz $U(\boldsymbol{\theta})$, composed of 5 qubits
with 8 layers. We use the ansatz in the numerical demonstration in Section~\ref{SECTION-experiment}.
Each layer is composed of the set of parameterized single-qubit rotational
gate $R_y(\theta_r)=\exp(-i \theta_r \sigma_y /2)$ and CNOT gates that
connect adjacent qubits;
$\theta_r$ is the $r$-th parameter and $\sigma_y$ is the Pauli $Y$ operator (hence $U(\boldsymbol{\theta})$ is a real matrix).
We randomly initialize all $\theta_r$ at the beginning of each training.}{FIGURE-data-circuit}{width=520pt}

\subsection{The proposed algorithm}

This section is twofold; first we identify a condition that guarantees the
equality in Eq.~(\ref{EQUATION-goal}); then, based on this condition, we specify
a valid cost function and describe the design procedure of $U(\boldsymbol{\theta})$.
The algorithm depends on the following two cases related to the elements of
target $\dvector$:
\begin{description}
	\item[(Case 1)]
	      The elements of $\dvector$ are all non-positive or all non-negative.
	\item[(Case 2)]
	      Otherwise.
\end{description}

It should be noted that, even in Case 1, the previously proposed method \cite{zoufal2019quantum}
does not always load the signs correctly.
For instance, suppose that we aim to create the ansatz state to approximate the target data state
$\data=(|0\rangle + |1\rangle + |2\rangle + |3\rangle)/2$.
The method \cite{zoufal2019quantum} only guarantees that, even ideally (i.e., the case where the
cost takes the minimum value zero), the absolute value of the amplitude of
$U(\boldsymbol{\theta}^*)|0\rangle^{\otimes n}$ is $(1/2, 1/2, 1/2, 1/2)$; but the output state can be e.g.,
$U(\boldsymbol{\theta}^*)|0\rangle^{\otimes n}=(|0\rangle - |1\rangle + |2\rangle - |3\rangle)/2$.
On the other hand, our method guarantees that, in the ideal case, the output
state is exactly the target state, i.e.,
$U(\boldsymbol{\theta}^*)|0\rangle^{\otimes n}=(|0\rangle + |1\rangle + |2\rangle + |3\rangle)/2=\data$.

\subsubsection{Condition for the perfect encoding}

In Case 1, we consider the following two conditions:
\begin{align}
	\label{EQUATION-caseone-condition-1}
	|\langle j|U(\boldsymbol{\theta})|0\rangle^{\otimes n}|^2 & = \dvector_j^2 \qquad (\forall j)
	\\
	\label{EQUATION-caseone-condition-2}
	|\langle j|H^{\otimes n}U(\boldsymbol{\theta})|0\rangle^{\otimes n}|^2
	                                                    & = \left(\sum_{k=0}^{N-1}\dvector_k\langle j|H^{\otimes n}|k\rangle\right)^2 \\
	                                                    & \equiv \left(\dvector_j^{H}\right)^2 \qquad (\forall j)
	\nonumber
\end{align}
Note that $\dvector_j^{H}$ is classically computable with complexity $O(N\log N)$,
by using the Walsh-Hadamard transform \cite{ahmed1975walsh};
in particular, if $\dvector$ is a sparse vector, this complexity can be reduced;
that is, if $\dvector$ has only $K=N^{\alpha}$ non-zero elements $(0<\alpha<1)$, there exists
a modified Walsh-Hadamard-based algorithm with computational complexity $O(K\log K\log (N/K))$
such that the success probability asymptotically approaches to $1$ as $N$
increases \cite{scheibler2015afast}.

If both two conditions \eqref{EQUATION-caseone-condition-1} and \eqref{EQUATION-caseone-condition-2} are satisfied, it is guaranteed that our goal is exactly satisfied, which is stated in the
following theorem (the proof is found in Appendix \ref{APPENDIX-proof}):

\begin{th.}
\label{THEOREM-validity}
In Case 1, if the $n$-qubits PQC $U(\boldsymbol{\theta})$ satisfies
Eqs.~(\ref{EQUATION-caseone-condition-1}) and
(\ref{EQUATION-caseone-condition-2}), then
$ U(\boldsymbol{\theta})|0\rangle^{\otimes n} = \sum_j \dvector_j |j\rangle$ or
$ U(\boldsymbol{\theta})|0\rangle^{\otimes n} = -\sum_j \dvector_j |j\rangle$ holds.
\end{th.}

\fighalf{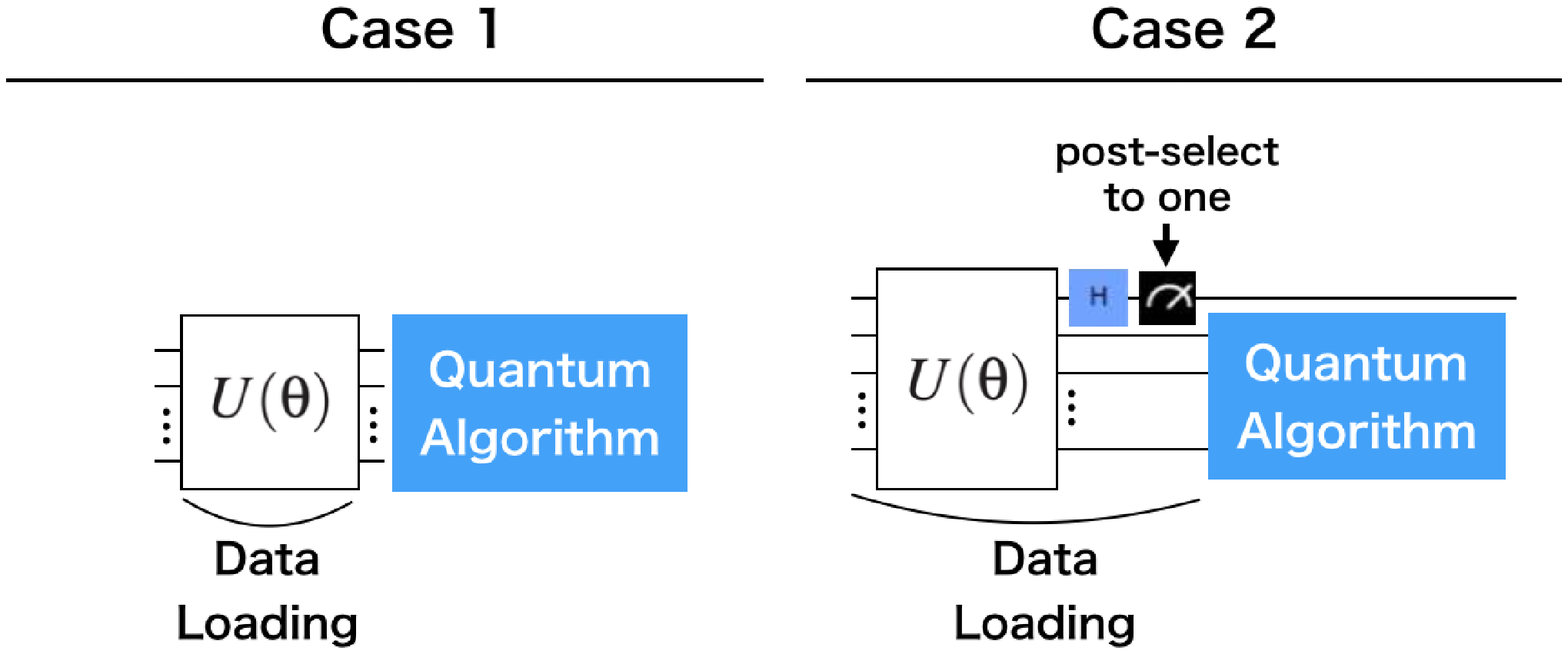}
{Overview of the data loading in Case 1 and Case 2.  }
{FIGURE-algorithm-overview}{width=250pt}

In Case 2, i.e., the case where some (not all) elements of $\dvector$ are
non-negative while the others are positive, the target state $\data$
can be decomposed to
\begin{equation}
	\begin{split}
		\data &= |Data^{+}\rangle + |Data^-\rangle \\
	\end{split},
\end{equation}
where the amplitudes of $|Data^{+}\rangle$ are positive and those of
$|Data^{-}\rangle$ are non-positive.
Then, by introducing an auxiliary single qubit, we can represent the
state $\data$ in the form considered in Case 1; that is, the amplitudes of the
$(n+1)$-qubits state
\begin{equation}
	\label{EQUATION-sign-manipulated-state}
	|\Bar{\psi}\rangle
	\equiv |Data^{+}\rangle|0\rangle - |Data^{-}\rangle|1\rangle
\end{equation}
are non-negative and $\langle\Bar{\psi}|\Bar{\psi}\rangle = 1$.
We write this state as
$|\Bar{\psi}\rangle = \sum_{i=0}^{2N-1}\bold{\Bar{d}}_j|j\rangle$ in terms of
the computational basis $\{|j\rangle\}$ and the corresponding $2N$-dimensional
vector $\bold{\Bar{d}}$.
Then, Theorem \ref{THEOREM-validity} states that, if the condition
\begin{align}
	\label{EQUATION-casetwo-condition-1}
	|\langle j|U(\boldsymbol{\theta})|0\rangle^{\otimes n+1}|^2                & = \Bar{\dvector}_j^2\qquad (\forall j)
	\\
	\label{EQUATION-casetwo-condition-2}
	|\langle j|H^{\otimes n+1}U(\boldsymbol{\theta})|0\rangle^{\otimes n+1}|^2 & = \left(\sum_{k=0}^{2N-1}\Bar{\dvector}_k\langle j|H^{\otimes n+1}|k\rangle\right)^2 \\
	                                                                     & \equiv \left(\Bar{\dvector}_j^{H}\right)^2\qquad (\forall j)
	\nonumber
\end{align}
are satisfied, then
$U(\boldsymbol{\theta})|0\rangle^{\otimes n+1} = \pm |\Bar{\psi}\rangle$ holds.
Further, once we obtain $|\Bar{\psi}\rangle$, this gives us the target $\data$,
via the following procedure.
That is, operating the Hadamard transform to the last auxiliary qubit yields
\begin{equation}
	\label{EQUATION-postselection-target}
	\begin{split}
		I^{\otimes n}\otimes H|\Bar{\psi}\rangle &= \frac{|Data^{+}\rangle - |Data^{-}\rangle}{\sqrt{2}} |0\rangle
		\\&+ \frac{|Data^{+}\rangle + |Data^{-}\rangle}{\sqrt{2}}|1\rangle,
	\end{split}
\end{equation}
and then the post-selection of $|1\rangle$ via the measurement on the last qubit
in Eq.~(\ref{EQUATION-postselection-target}) gives us $\data$ in the first
$n$-qubits.
The above result is summarized as follows.

\begin{th.}
\label{THEOREM-validity-2}
In Case 2, suppose that the $(n+1)$-qubits PQC $U(\boldsymbol{\theta})$ satisfies
Eqs.~(\ref{EQUATION-casetwo-condition-1}) and
(\ref{EQUATION-casetwo-condition-2}).
Then, if the measurement result of the last qubit in the computational basis
for the state $(I^{\otimes n}\otimes H)U(\boldsymbol{\theta})|0\rangle^{\otimes n+1}$
is $|1\rangle$, then $\data$ is generated.
That is,
\begin{equation*}
	(I^{\otimes n}\otimes |1\rangle \langle 1|)(I^{\otimes n}\otimes H)U(\boldsymbol{\theta})|0\rangle^{\otimes n+1} \propto \data |1\rangle.
\end{equation*}
\end{th.}

This theorem implies that, if $U(\boldsymbol{\theta})$ is trained so 
that the conditions (\ref{EQUATION-casetwo-condition-1}) and 
(\ref{EQUATION-casetwo-condition-2}) are satisfied,  $\data$ can be 
obtained by the above post-selection procedure, with success probability 
nearly $1/2$. 
Note that, by applying the extra post processing, we can obtain $\data$ 
with success probability 1 instead of $1/2$, which is shown in 
Appendix \ref{SECTION-amplification}. 
The overview of the data-loading circuits in Case 1 and Case 2 are 
summarized in Fig.~\ref{FIGURE-algorithm-overview}. 
As seen from the figure, $U(\boldsymbol{\theta})$ is directly used as 
the data loading circuit in Case 1, while we need the post-processing 
after $U(\boldsymbol{\theta})$ in Case 2.

\subsubsection{Optimization of $U(\boldsymbol{\theta})$}

Here we provide a training method for optimizing $U(\boldsymbol{\theta})$, so that
Eqs. (\ref{EQUATION-caseone-condition-1}) and (\ref{EQUATION-caseone-condition-2})
are nearly satisfied in Case 1, and Eqs.~(\ref{EQUATION-casetwo-condition-1}) and
(\ref{EQUATION-casetwo-condition-2}) are nearly satisfied in Case 2, with as small
approximation error as possible.
For this purpose, we employ the strategy to decrease the {\it maximum mean
		discrepancy (MMD)} cost \cite{Sriperumbudur2008InjectiveHS,Fukumizu2007KernelMO}, which was previously proposed for training Quantum Born Machine \cite{liu2018differentiable,coyle2020born}.
Note that the other costs, the Stein discrepancy (SD) \cite{coyle2020born} and the
Sinkhorn divergence (SHD) \cite{coyle2020born} can also be taken, but in this paper we
use MMD for its ease of use.

The MMD is a cost of the discrepancy between two probability distributions:
$q_{\theta}(j)$, the model probability distribution, and $p(j)$, the target
distribution.
The cost function $\mathcal{L}_{MMD}(q_{\theta}, p)$ is defined as
\begin{equation}
	\label{EQUATION-mmd-definition}
	\begin{split}
		\mathcal{L}_{MMD}(q_{\theta}, p) &\equiv \gamma_{ MMD}(q_{\theta}, p)^2, \\
		\gamma_{MMD}(q_{\theta}, p) &= \left|\sum_{j=0}^{N-1} q_\theta(j)\bold{\Phi}(j) - \sum_{j=0}^{N-1} p(j)\bold{\Phi}(j)\right|,
	\end{split}
\end{equation}
where $\bold{\Phi}(j)$ is a function that maps $j$ to a feature space.
Thus, given the kernel $\kappa(j, k)$ as
$\kappa(j, k)=\bold{\Phi}(j)^{T}\bold{\Phi}(k)$, it holds
\begin{equation}
\label{EQUATION-mmd-expectation}
	\begin{split}
		\mathcal{L}_{MMD}(q_{\theta}, p) &= \E{\substack{j\sim q_{\theta}\\k\sim q_{\theta}}}{\kappa(j,k)}
		-2 \E{\substack{j\sim q_{\theta}\\k\sim p}}{\kappa(j,k)} \\
		& ~~ + \E{\substack{j\sim p\\k\sim p}}{\kappa(j,k)},
	\end{split}
\end{equation}
where, for example one of the expectation values is defined by
\begin{equation}
\label{EQUATION-MMD-term}
     \E{\substack{j\sim q_{\theta}\\k\sim q_{\theta}}}{\kappa(j,k)} = \sum_{j=0}^{N-1}\sum_{k=0}^{N-1}\kappa(j,k)q_{\theta}(j)q_{\theta}(k).
\end{equation}

Note that, even though the index of the sum goes till $N-1$ in Eq.~\eqref{EQUATION-MMD-term}, we can efficiently estimate the expectation 
value without $O(N)$ computation by sample-averaging as follows. 
First, given $N_{\rm shot}$ as the number of samples for each index, we sample 
$\{j_{\ell}\}_{\ell=0}^{N_{\rm shot}-1}$ and $\{k_{\ell}\}_{\ell=0}^{N_{\rm shot}-1}$ 
according to the probability distribution $q_{\theta}(\cdot)$; note that, in our case, 
$q_{\theta}(j)$ is the probability to obtain $|j\rangle$ as a result of measuring 
the final state of PQC, and thus we can obtain samples just by measuring the final 
state multiple times. 
Then, using the samples $\{j_{\ell}\}_{\ell=0}^{N_{\rm shot}-1}$ and 
$\{k_{\ell}\}_{\ell=0}^{N_{\rm shot}-1}$, we can approximate the expectation value as
\begin{align}
\label{EQUATION-mmd-sample}
     {\mathbf E}_{\substack{j\sim q_{\theta}\\k\sim q_{\theta}}}[\kappa(j,k)] 
     \simeq \frac{1}{N_{\rm shot}}\sum_{\ell=0}^{N_{\rm shot}-1}\kappa(j_{\ell}, k_{\ell}). 
\end{align}
The approximation error is bounded by $O(1/\sqrt{N_{\rm shot}})$ with high probability; 
this fact can be proven by using the bound for probability distributions such as Chernoff 
bound \cite{chernoff1952measure} combined with the technique to derive the error bound, 
e.g. \cite{Sriperumbudur2009OnIP,coyle2020born}. 
Similarly, we can efficiently estimate the other expectation values in $\mathcal{L}_{MMD}$ 
by the sample-averaging technique; as a result, we can estimate $\mathcal{L}_{MMD}$ with 
guaranteed error $O(1/\sqrt{N_{\rm shot}})$, via $O(N_{\rm shot})<O(N)$ computation.

It should also be noted that when the kernel is {\it characteristic}
\cite{Sriperumbudur2008InjectiveHS,Fukumizu2007KernelMO}, then
$\mathcal{L}_{MMD}(q_{\theta}, p)=0$ means $q_{\theta}(j)=p(j)$ for all $j$.
In this paper, we take one dimensional Gaussian kernel $\kappa(j, k)=C\exp(-(j-k)^2/2\sigma^2)$ with a positive constant $C$,  which is characteristic. 

In Case 1, the goal is to train the model distributions
\begin{eqnarray*}
	q_{\theta}(j) &= |\langle j|U(\boldsymbol{\theta})|0\rangle^{\otimes n}|^2,
	\\
	q_{\theta}^H(j) &= |\langle j|H^{\otimes n}U(\boldsymbol{\theta})|0\rangle^{\otimes n}|^2
\end{eqnarray*}
so that they approximate the target distributions
\begin{equation}
	p(j) = \dvector_j^2, ~~ p^H(j) = \left( \dvector_j^{H}\right)^2,
\end{equation}
respectively.
In Case 2, the model distributions
\begin{eqnarray*}
	q_{\theta}(j) &= |\langle j|U(\boldsymbol{\theta})|0\rangle^{\otimes n+1}|^2,\\
	q_{\theta}^H(j)&= |\langle j|H^{\otimes n+1}U(\boldsymbol{\theta})|0\rangle^{\otimes n+1}|^2
\end{eqnarray*}
are trained so that they approximate the target distributions
\begin{equation}
	p(j) = \Bar{\dvector}_j^2, ~~ p^H(j) = \Bar{\dvector}_j^{H2},
\end{equation}
respectively.
In both cases, our training policy is to minimize the following cost
function:
\begin{equation}
	\label{EQUATION-cost-function}
	\mathcal{L}(\boldsymbol{\theta}) = \frac{\mathcal{L}_{MMD}(q_{\theta}, p) + \mathcal{L}_{MMD}(q^{H}_{\theta}, p^H)}2.
\end{equation}
Actually, $\mathcal{L}(\boldsymbol{\theta})$ becomes zero if and only if
$\mathcal{L}_{MMD}(q_{\theta}, p) = 0$ and
$\mathcal{L}_{MMD}(q^{H}_{\theta}, p^H) = 0$, or equivalently
$q_{\theta}(j) = p(j)$ and $q^{H}_{\theta}(j) = p^H(j)$ for all $j$ as long as we use a characteristic kernel.

To minimize the cost function \eqref{EQUATION-cost-function}, we take the
standard gradient descent algorithm.
In particular as we note at the end of Section~\ref{SECTION-goal}, we consider the PQC where each parameter $\theta_r$ is embedded
in the quantum circuit in the form $\exp(- i \theta_r \sigma_y/2)$.
In this case, the gradients of $q_\theta$ and $q_\theta^H$ with respect to $\theta_r$ can be
computed by using the parameter shift rule \cite{Crooks2019GradientsOP} as
\begin{equation}
\begin{split}
\label{eq:shift-rule}
    \frac{\partial q_\theta(j)}{\partial \theta_r} &= q_{{\theta}_r}^{+}(j) - q_{{\theta}_r}^{-}(j), \\
    \frac{\partial q^H_\theta(j)}{\partial \theta_r} &= q_{{\theta}_r}^{H+}(j) - q_{{\theta}_r}^{H-}(j),
\end{split}
\end{equation}
where $q_{\theta_r}^{\pm}(j) = |\langle j|U_{r\pm}(\boldsymbol{\theta})|0\rangle|^2$, $q_{\theta _r}^{H\pm}(j) = |\langle j|HU_{r\pm}(\boldsymbol{\theta})|0\rangle|^2$. The shifted unitary operator is defined by
\begin{equation}
\label{EQUATION-parameter-shift}
	\begin{split}
		U_{r \pm}(\boldsymbol{\theta})
		&=U_{r \pm}(\{\theta_1, \cdots, \theta_{r-1}, \theta_r, \theta_{r+1}, \cdots, \theta_R\})
		\\
		&= U(\{\theta_1, \cdots, \theta_{r-1}, \theta_r\pm\pi/2, \theta_{r+1}, \cdots, \theta_R\}),
	\end{split}
\end{equation}
with $R$ as the number of the parameters, which can be written as $R=\ell n$ (recall that $\ell$ is the depth of PQC). Then, by differentiating  \eqref{EQUATION-mmd-expectation} and using \eqref{eq:shift-rule}, the gradient of $\mathcal{L}$ can be explicitly computed  \cite{liu2018differentiable} as%
\begin{equation}
	\label{EQUATION-cost-gradient}
	\begin{split}
		2\frac{\partial\mathcal{L}}{\partial\theta_r}
		&=
		\E{
			\substack{j\sim q_{\theta_r}^{+}\\k\sim q_{\theta}}
		}{\kappa(j, k)}
		- \E{
			\substack{j\sim q_{\theta_r}^{-}\\k\sim q_{\theta}}
		}{\kappa(j, k)} \\
		&- \E{
			\substack{j\sim q_{\theta_r}^{+}\\k\sim p}
		}{\kappa(j, k)}
		+ \E{
			\substack{j\sim q_{\theta_r}^{-}\\k\sim p}
		}{\kappa(j, k)} \\
		&+
		\E{
			\substack{j\sim q_{\theta_r}^{H+}\\k\sim q_{\theta}^H}
		}{\kappa(j, k)}
		- \E{
			\substack{j\sim q_{\theta_r}^{H-}\\k\sim q_{\theta}^H}
		}{\kappa(j, k)} \\
		&- \E{
			\substack{j\sim q_{\theta_r}^{H+}\\k\sim p^H}
		}{\kappa(j, k)}
		+ \E{
			\substack{j\sim q_{\theta_r}^{H-}\\k\sim p^H}
		}{\kappa(j, k)}.
	\end{split}
\end{equation}

We can approximately compute the gradient (\ref{EQUATION-cost-gradient}) by sampling 
$j$ and $k$ from the distributions $q_{\theta}$, $q_{\theta_r}^+$, $q_{\theta_r}^-$, 
$q_{\theta}^H$, $q_{\theta_r}^{H+}$, $q_{\theta_r}^{H-}$ $p$, and $p^H$ 
similar to the case of Eq.~$\eqref{EQUATION-mmd-sample}$. 
Then, using the gradient descent algorithm with Eq.~(\ref{EQUATION-cost-gradient}),
we can update the vector $\boldsymbol{\theta}=(\theta_1, \ldots, \theta_R)$ 
to the direction that minimizes $\mathcal{L}(\boldsymbol{\theta})$. 
Note that, in the above sampling approach, the estimation error of the 
gradient vectors does not depend on $N$ but only on the number of samples 
$N_{\rm shot}$. 
This is because each gradient is written as the sum of the expectation values 
\eqref{EQUATION-cost-gradient} and each expectation value can be estimated 
with the error $O(1/\sqrt{N_{\rm shot}})$, as discussed around 
Eq.~\eqref{EQUATION-mmd-sample}.

Lastly we remark that we may be able to utilize the classical shadow technique
\cite{Huang_2020,huang2021efficient} and its extension \cite{hillmich2021decision}, to significantly reduce the number of
measurements as follows.
Recall that the density matrix of an $n$-qubit quantum state $\rho$ is a linear combination of
$4^n$ Pauli terms written as $\rho = \sum_{P} \alpha_P P$ where $\alpha_P\in \mathbf{R}$ and
$P \in \{I, X, Y, Z\}^{\otimes n}$.
Also in the proposed method, the quantum state generated from the PQC is measured either
in the computational basis (i.e., the eigenstates of $Z^{\otimes n}$) or in the rotated computational
basis via the Hadamard gate (i.e., the eigenstates of $X^{\otimes n}$).
Applying the classical shadow technique allows us to estimate all coefficients $\alpha_P$ for $P \in \{I, Z\}^n$ and for $P \in \{I, X\}^n$ within an additive error $\epsilon$ by spending $\propto \mbox{poly}(n)/\epsilon^2$ number of measurements.
We can also extend the measurement basis by the eigenstates of $Y$,
and by applying the classical shadow technique to probabilistically check if the coefficients of
$\alpha_P$'s are correct. We leave the details of the analysis for future work.


\subsection{Computational complexity of the AAE algorithm}
\label{SECTION-computational-complexity}

We have seen above that the number of measurements required for estimating the cost and the
gradient vector does not scale exponentially with the number of qubits.
However, we must still solve the issues that often appear in the standard variational quantum
algorithms (VQAs), for ensuring the scalability of our algorithm.
Below we pose a few typical issues and describe how we could handle them.

Firstly, it is theoretically shown that the general VQA with highly-expressive PQC
has the barren plateau issue \cite{mcclean2018barren}; that is, the gradient of the
cost function becomes exponentially small as the number of qubits increases.
Our algorithm may also have the same issue, even though $U(\boldsymbol{\theta})$ is limited to
be real in the computational basis.
For mitigating this issue, several approaches have been proposed; e.g., circuit
initialization \cite{Grant2019initialization}, special structured ansatz
\cite{cerezo2021cost}, and parameter embedding \cite{volkoff2021large}, while
further studies are necessary to examine the validity of these approaches.
The methods \cite{Grant2019initialization} and \cite{volkoff2021large} are
applicable to our algorithm; on the other hand, the approach of \cite{cerezo2021cost}
depends on the detail of the cost function (the MMD cost in our case), and we need further investigation to see the applicability of this method to our algorithm.
Related to this point, we also need to carefully address the issue that
the landscape of the cost function in VQA may have many local minima, which may
result in many trials of the training for PQC.
Because this local minima issue is ubiquitous in classical optimization problems,
we can employ several established classical optimizers \cite{wierichs2020avoiding}.
Other approaches may also be applicable to our algorithm, but their theoretical understanding, particularly the convergence proof, are still to be investigated.

The next typical issue that our algorithm shares in common with the general
VQA is that the depth of circuit tends to become bigger in order to reduce
the value of cost function below a certain specified value \cite{cerezo2020variational}.
Various attempts to reduce the circuit depth have been made in the literature
\cite{grimsley2019adaptive,tang2021qubit,ryabinkin2018qubit,tkachenko2021correlation} and some of those are applicable to our algorithm.
Also, Refs.~\cite{tang2021cutqc,peng2020simulating} provide methods to split
a large quantum circuit to several small quantum circuits.
Even with those assistance, we need further theoretical development to conclude that $O(\mbox{poly}(n))$ depth is actually enough for training any data loading circuit.
Nevertheless, in contrast to the problems that require near-perfect cost minimization such as VQE in chemistry, the depth needed for our algorithm is smaller as long as very precise data loading is not necessary.
Computation of the singular value decomposition entropy, which we will discuss in Section~\ref{SECTION-experiment}, is an example of such quantum algorithms.

\begin{table*}[ht]
\centering
\caption{
Overview of the computational complexity for our algorithm (AAE) and the
case for exact encoding \cite{plesch2011quantum,shende2006synthesis,grover2002creating,Mttnen2005TransformationOQ,Shende2005QuantumCF,A1,A3,A4,A6,Bergholm_2005,A10,B1,malvetti2021quantum} ; for each case we show both the results when the data vector is dense or sparse. For exact encoding with sparse data, we show the result in Ref.~\cite{malvetti2021quantum}. 
We divide the computational complexity into two stages: the training stage
(training) and the execution stage of the quantum algorithm (execution).
In the case of exact encoding, there is no training stage.
The total number of gate operations required in the training stage of AAE
is denoted by $N_{train}$ (as the gate operations, we include both the single qubit operations and the two-qubit operations).
The number of gates required for the main quantum algorithm is denoted by $N_{alg}$, and the total number of measurements required to retrieve the
output satisfying a sufficient precision for each problem is denoted by $N_{mes}$.
}
\scalebox{1}{
\begin{tabular}{|l|l|c|c|c|c|c|}
\hline
\multicolumn{3}{|l|}{\multirow{2}{*}{strategy}}
   &
  \multicolumn{2}{c|}{(1) AAE} &
  \multicolumn{2}{c|}{\begin{tabular}[c]{@{}c@{}}(2) exact encoding \end{tabular}} \\ \cline{4-7}
  \multicolumn{3}{|c|}{}&dense& sparse & dense & sparse \\ \hline
\multicolumn{3}{|l|}{\# of nonzero elements in the data}   &
  \hspace{0.6cm}$N$\hspace{0.6cm} &
  $K$ &
  $N$ &
  $K$ \\ \hline
\multicolumn{3}{|l|}{\# of gates in the data loading circuit}
   &
  \multicolumn{2}{c|}{$O(\mbox{poly}(\log N))$} &
  $O(N)$ &
  $O(K\log N)$ \\ \hline \hline
\multirow{3}{*}{\begin{tabular}[c]{@{}l@{}}computational\\ complexity\\ (training)\end{tabular}} &
 \multicolumn{2}{|l|}{
 \begin{tabular}[c]{@{}l@{}l}classical\\  (Walsh Hadamard Transform)\end{tabular}}
   &
  $\ \ O(N\log N)\ \ $&
  \begin{tabular}[c]{@{}c@{}}$\ \ O(K\log K\ \  $\\$\times\log(\frac{N}{K}))$ \end{tabular}
   &
  \multicolumn{2}{c|}{\multirow{3}{*}{-}} \\ \cline{2-5}
 &
 \multicolumn{2}{l|}{
 \begin{tabular}[c]{@{}l@{}l}quantum\\ (total \# of gate operations)\end{tabular}
  } &
\multicolumn{2}{c|}{$N_{train}$} &

  \multicolumn{2}{c|}{} \\ \hline \hline
\multirow{5}{*}{\begin{tabular}[c]{@{}l@{}}computational\\complexity\\ (execution)\end{tabular}} &
  \multicolumn{2}{l|}{\begin{tabular}[c]{@{}l@{}}(a) \# of gate operations for \\ the data loading per\\ one measurement\end{tabular}} &
  \multicolumn{2}{c|}{$O(\mbox{poly}(\log N))$} &
  $O(N)$ &
  $O(K)$ \\ \cline{2-7}
 &
  \multicolumn{2}{l|}{\begin{tabular}[c]{@{}l@{}}(b) \# of gate operations for\\ quantum algorithm\\ per one measurement\end{tabular}} &
  \multicolumn{4}{c|}{$N_{alg}$} \\ \cline{2-7}
 &
  \multicolumn{2}{l|}{\begin{tabular}[c]{@{}l@{}}(c) \# of measurements\end{tabular}} &
  \multicolumn{4}{c|}{$N_{mes}$} \\ \cline{2-7}
 &
  \multicolumn{2}{l|}{total = $[(a) + (b)]\times(c)$} &
  \multicolumn{2}{c|}{\begin{tabular}[c]{@{}c@{}}$\bm{O((\mbox{poly}(\log N)+N_{alg})}$\\ $\bm{\times N_{mes}}$\end{tabular}} &
  \begin{tabular}[c]{@{}c@{}}$\bm{O(N + N_{alg})}$\\ $\bm{\times N_{mes}}$\end{tabular} &
  \begin{tabular}[c]{@{}c@{}} $\bm{O(K + N_{alg})}$\\ $\bm{\times N_{mes}}$\end{tabular} \\ \hline
\end{tabular}
}
\label{TABLE-complexity}
\end{table*}

TABLE \ref{TABLE-complexity} summarizes the computational complexity of AAE,  under the assumption that the PQC with the number of gates $O(\mbox{poly}(n))$
achieves the approximate data loading with sufficient precision for the given problem. We show two cases: (1) the case when using AAE (denoted by ``AAE" in the table) and (2) the case when exactly encoding data  (denoted by ``exact encoding") \cite{plesch2011quantum,shende2006synthesis,grover2002creating,Mttnen2005TransformationOQ,Shende2005QuantumCF,A1,A3,A4,A6,Bergholm_2005,A10,B1,malvetti2021quantum} ; for each case we show both the results when the data vector is dense or sparse. For exact encoding with sparse data, we show the result in Ref.~\cite{malvetti2021quantum}. Also, the number of non-zero elements in sparse data is denoted by $K$.
We divide the computational complexity into two stages: the training stage  (training) and the execution stage of the quantum algorithm (execution).
In the case of exact encoding, there is no training stage.
The total number of gate operations required in the training stage of AAE
is denoted by $N_{train}$, and that for the main quantum algorithm is
denoted by $N_{alg}$. 
The total number of measurements required to retrieve the output of the algorithm, satisfying sufficient precision for the given problem, is denoted by $N_{mes}$.
We emphasize the total computational complexity in the execution stage by bold letters. 

Regarding the computational complexity, the merit of using AAE exists 
in the computational complexity in execution stage. 
In particular, the computational complexity of AAE is $O(N)$ times smaller
in the execution stage than that of the exact encoding method, when 
$N_{alg}$ is of the order of $\mbox{poly}(\log N)$. 
Such a merit is favorable in particular when the data loading circuit is 
used repeatedly. 
One example is when $N_{mes}$ is as large as $\mbox{poly}(\log N)$. 
Another example is when the data loading circuit is usable in various  
problems; for example, once we load a training dataset for a particular 
quantum machine learning problem, it can be utilized in other machine 
learning models.

For the training stage of AAE, however, we need further discussion.
Firstly, the $O(N\log N)$ classical computation for the case of dense data,
which comes from the Walsh-Hadamard transform in AAE, seems costly.
However, other data loading methods implicitly contains processes to
register the data.
For instance, the exact encoding method shown in the table needs the process that
compiles the data into $O(N)$ quantum gates, which at least requires $O(N)$ computational complexity.
Also, even when a quantum random access memory (QRAM) \cite{qram} is ideally
available, registering $O(N)$ data into the QRAM requires at least $O(N)$
computational complexity; e.g., in the QRAM proposed in Ref.~\cite{park2019circuit}, $O(N\log N)$ gate operations are necessary for the data registration.

Secondly, related with the above-mentioned trainability issues in VQA,
$N_{train}$ might become large, in the absence of some elaborated techniques for
VQA.
A promising approach is to take the convex relaxation on the target cost function,
in which case the total number of iterations to achieve $\tilde{\mathcal{L}}(\boldsymbol{\theta})<\epsilon$ is $O(\mbox{poly}(\log N)/\epsilon^2)$ \cite{Harrow2019LowdepthGM}, where $\tilde{\mathcal{L}}$ is the relaxed convex
cost function.
Furthermore, it was shown in \cite{Harrow2019LowdepthGM} that the total number of measurements for realizing $\tilde{\mathcal{L}}(\boldsymbol{\theta}) < \epsilon$ is $O(GS/\epsilon^2)$,
where $G$ is the upper bound of $|\partial \tilde{\mathcal{L}}/\partial \theta_r|$
over the parameter space and $S$ is a constant that relates with the size of the parameter space.
Thus, combining these two complexities, we find that the total number of
gate operations $N_{train}$ to achieve $\tilde{\mathcal{L}}(\boldsymbol{\theta}) < \epsilon$
is $O(\mbox{poly}(\log N)\mbox{poly}(1/\epsilon))$, provided that the number of
parameters is of the order of $\mbox{poly}(\log N)$.
Note that, of course, the convex relaxation appends an additional error related
with the gap between the original cost function and the relaxed one.
Nonetheless, we hope that, in our case, this gap could be minor compared to the
target precision of the cost function (SVD entropy in our case);
we will study this problem as an important future work for having scalability.


\subsection{Some modification on the AAE algorithm}

Before concluding this section, we consider four types of modifications 
on the AAE algorithm. 
The first two are the change of the cost function, and the next one 
discusses the change of the conditions for perfect-encoding; 
the fourth one is on the possibility to formulate the AAE algorithm 
in the GAN framework.

The first one is simple; we may be able to build the cost function as 
the weighted average of $\mathcal{L}_{MMD}(q_{\theta}, p)$ and 
$\mathcal{L}_{MMD}(q^{H}_{\theta}, p^H)$ instead of the current 
equally-weighted average \eqref{EQUATION-cost-function}. 
It is worth investigating the effect of this modification.

The second possible change is taking a cost function other than MMD.
That is, as mentioned above, Stein discrepancy (SD) or Sinkhorn
divergence (SHD) can serve as a cost for measuring the difference
of two probability distributions. 
Also, as another type of cost function, readers may wonder if the Kullback–Leibler divergence (KL-divergence)
\begin{equation}
	\label{EQUATION-kl-divergence}
	\mathcal{L}_{\rm KL}(p, q_\theta) = \sum_{j=0}^{N-1} \Big[p(j)\log(p(j))
		- p(j)\log(q_\theta(j))\Big]
\end{equation}
would be a more natural cost function for comparing a target
distribution $p(j)$ and a model distribution $q_\theta(j)$ with parameter
$\theta$.

The gradient $\partial\mathcal{L}_{\rm KL}/\partial\theta_r$ is given by
\begin{align}
	\label{EQUATION-kl-gradient}
	\frac{\partial\mathcal{L}_{\rm KL}}{\partial\theta_r}
	&= - \sum_{j=0}^{N-1} \frac{p(j)}{q_\theta(j)}
	\frac{\partial q_\theta(j)}{\partial \theta_r} \nonumber \\
	&= - \sum_{j=0}^{N-1} \frac{p(j)}{q_\theta(j)}
	(q_{\theta_{r}}^+(j) - q_{\theta_{r}}^-(j)) \nonumber \\
	&= -{\mathbf E}_{
			j\sim q_{\theta_r}^{+}} \left[\frac{p(j)}{q_\theta(j)}\right]
			+ {\mathbf E}_{
			j\sim q_{\theta_r}^{-}}\left[\frac{p(j)}{q_\theta(j)}\right]. 
\end{align}
However, we cannot efficiently compute this quantity by sample-averaging 
unlike the case of MMD. 
For example, we sample $\{j_{\ell}\}_{\ell=0}^{N_{shot}-1}$ from 
$q_{\theta_r}^{+}(\cdot)$ and may compute the first term of the gradient 
as
\begin{equation}
    {\mathbf E}_{
			j\sim q_{\theta_r}^{+}} \left[\frac{p(j)}{q_\theta(j)}\right] \simeq \frac{1}{N_{\rm shot}}\sum_{\ell=0}^{N_{\rm shot} -1}\frac{p(j_{\ell})}{q_{\theta}(j_{\ell})},
\end{equation}
similar to the case of Eq.~\eqref{EQUATION-mmd-sample}. 
Then we need to compute the value of 
$q_{\theta}(j) = |\langle j|U({\boldsymbol \theta})|0\rangle^{\otimes n}|^2$ 
for each $\ell$, which however requires $O(2^n)$ measurements. 
Therefore, the gradient of the KL-divergence cannot be efficiently computed 
in our setting, unlike the case of MMD (see \cite{liu2018differentiable} 
for more detailed explanation). 

To the contrary, the gradient of SD and SHD as well as MMD are 
efficiently computable, because the gradient vector is written in terms 
of the averages of efficiently computable statistical quantities as in 
Eq.~\eqref{EQUATION-cost-gradient} \cite{coyle2020born}.

The third possible change is altering the conditions  \eqref{EQUATION-caseone-condition-2} and \eqref{EQUATION-casetwo-condition-2},
which is used for characterizing the perfect encoding.
In Case 1, we train $U(\boldsymbol{\theta})$ so that Eqs.~(\ref{EQUATION-caseone-condition-1}) and (\ref{EQUATION-caseone-condition-2}) are approximately satisfied; the complexity for computing the right hand side of Eq.~(\ref{EQUATION-caseone-condition-2}) is $O(N{\rm log}N)$.
However, as seen in the proof of Theorem \ref{THEOREM-validity}, even if the condition (\ref{EQUATION-caseone-condition-2}) is replaced by
\begin{equation}
	\label{EQUATION-another-caseone-condition-2}
	|\langle 0|H^{\otimes n}U(\boldsymbol{\theta})|0\rangle^{\otimes n}|^2 = \left(\sum_{k=0}^{N-1}\dvector_k\langle 0|H^{\otimes n}|k\rangle\right)^2,
\end{equation}
the perfect encoding is still achieved; that is,
$U(\boldsymbol{\theta})|0\rangle = \sum_j \dvector_j |j\rangle$ or
$U(\boldsymbol{\theta})|0\rangle = -\sum_j \dvector_j |j\rangle$ holds.
This implies that we can obtain the data loading circuit by training
$U(\boldsymbol{\theta})$ so that Eqs.~(\ref{EQUATION-caseone-condition-1}) and (\ref{EQUATION-another-caseone-condition-2}) are approximately satisfied.
Then the complexity for computing the right hand side is reduced to $O(N)$.
In Case~2, the situation is the same.
Therefore, the modified algorithm with the use of the conditions (\ref{EQUATION-caseone-condition-1}) and (\ref{EQUATION-another-caseone-condition-2}) may also work.

Now, as another possibility of changing the conditions \eqref{EQUATION-caseone-condition-2} and \eqref{EQUATION-casetwo-condition-2},  readers may wonder that, if we carefully choose an operator $X$ instead of $H^{\otimes n}$, the conditions
\begin{align}
	\label{EQUATION-x-condition-1}
	|\langle j|U(\boldsymbol{\theta})|0\rangle^{\otimes n}|^2  & = \dvector_j^2,
	\\
	\label{EQUATION-x-condition-2}
	|\langle j|XU(\boldsymbol{\theta})|0\rangle^{\otimes n}|^2 & = \left(\sum_{k=0}^{N-1}\dvector_k\langle j|X|k\rangle\right)^2
\end{align}
would also result in $U(\boldsymbol{\theta})|0\rangle = \sum_j \dvector_j |j\rangle$ or $U(\boldsymbol{\theta})|0\rangle = -\sum_j \dvector_j |j\rangle$ for arbitrary real vector $\dvector$.
This is clearly favorable because we do not need Case~2; namely, we need
neither auxiliary qubits nor the post-selection.
However, as shown in Appendix \ref{APPENDIX-study}, it seems to be difficult to find such $X$ for arbitrary $\dvector$.
This is why we consider the two cases depending on $\dvector$.

The final possible change is utilizing GAN \cite{gan}, which was originally
proposed as the method to train a generative model.
GAN consists of two components: a generator and a discriminator.
The generator generates samples (fake data) and the discriminator receives
either a real data from a data source or fake data from the generator.
The discriminator is trained so that it exclusively classifies the fake data
as fake and the real data as real.
The generator is trained so that the samples generated by the generator are
classified as real by the discriminator.
If the training is successfully conducted, we will have a good generative model;
i.e., the probability distribution that governs the samples of the generator
well approximates the source distribution.
As mentioned in Section~I, the motivating work \cite{zoufal2019quantum} applied
GAN composed of the quantum generator implemented by the PQC and the classical
(neural network) discriminator, and demonstrated that the trained PQC approximates
the probability distribution $p(j)=\dvector_j^2$.
In our work, in contrast, we do not take the GAN formulation.
The main reason is that, in our case, the PQC is trained to learn {\it two}
probability distributions (Eqs.~\eqref{EQUATION-caseone-condition-1} and \eqref{EQUATION-caseone-condition-2} in Case 1), which cannot be formulated
in the ordinary GAN that handles only one generator and one discriminator.

However, customizing GAN to fit into our setting may be doable as follows;
we will discuss only Case 1, but the same argument applies to Case 2.
The customized GAN is composed of two quantum generators (Generator-A and
Generator-B) and two classical discriminators (Discriminator-A and
Discriminator-B).
In particular, we use one PQC to realize the two generators.
First, Generator-A and Discriminator-A correspond to the condition (4);
output samples of Generator-A (fake-data-A) are obtained by measuring the output
state of PQC in computational basis, and Discriminator-A receives the real data
sampled from $p(j)={\bf d}_j^2$ or the fake-data-A generated from Generator-A.
Also, Generator-B and Discriminator-B correspond to the condition (5); output
samples of Generator-B (fake-data-B) are generated by measuring the output state
of the same PQC yet in the Hadamard basis, and Discriminator-B receives the real
data sampled from $p^H(j)={\bf d}^{H2}_j$ or the fake-data-B generated by
Generator-B.
With this setting, the discriminators are trained so that they will exclusively
classify the real and fake data.
On the other hand, the PQC is trained so that the outputs of the generators are
to be classified as real by the discriminators.
Ideally, as a result of the training, we will obtain the generator that
almost satisfies \eqref{EQUATION-caseone-condition-1} and \eqref{EQUATION-caseone-condition-2}.


\section{Application to SVD entropy calculation for financial
  market indicator}
\label{SECTION-experiment}

This section is devoted to describe the quantum algorithm composed of our AAE and the
variational qSVD algorithm \cite{bravo2020quantum} for computing the SVD entropy for
stock price dynamics \cite{caraiani2014predictive}.
We first give the definition of SVD entropy and then describe the quantum algorithm, with
particular emphasis on how the AAE algorithm well fits into the problem of computing the
SVD entropy.
Finally the in-depth numerical simulation is provided.

\subsection{SVD Entropy}

The SVD entropy is used as one of the good indicators for forewarning the financial crisis, which
is computed by the singular value decomposition of the correlation matrix between stock prices.
Let $s_{j, t}$ be the price of the $j$-th stock at time $t$.
Then we define the logarithmic rate of return as follows;
\begin{equation}
	\label{EQUATION-logarithm}
	r_{jt} = \log(s_{j, t}) - \log(s_{j, t-1}).
\end{equation}
Also, the correlation matrix $C$ of the set of stocks $j=1,2,\ldots,N_s$ over the term $t=1,2,\ldots,T$
is defined as
\begin{equation}
	\label{EQUATION-covariance-matrix}
	C_{jk} = \sum_{t=1}^{T} a_{jt}a_{kt},
\end{equation}
where
\begin{equation}
	\label{EQUATION-stock-coefficient}
	a_{jt}=\frac{r_{jt}-\left\langle r_{j}\right\rangle}{\sigma_{j} \sqrt{N_s T}}.
\end{equation}
The average $\langle r_j \rangle$ and the standard deviation $\sigma_j$ over the whole period of term
are defined as
\begin{equation}
	\langle r_j\rangle = \frac{1}{T}\sum_{t=1}^{T} r_{jt}, ~~
	\sigma_j^2 = \frac{1}{T}\sum_{t=1}^{T}(r_{jt}-\langle r_j\rangle)^2.
\end{equation}
The correlation matrix $C$ is positive semi-definite, and thus its eigenvalues are non-negative. In addition, $C$ satisfies
%
\begin{equation}
	\label{C normalization}
	{\mathrm{Tr}}(C) =\sum_{j=1}^{N_s}\sum_{t=1}^T a_{jt}^2 = 1.
\end{equation}
Now for the positive eigenvalues $\lambda_1, \lambda_2, \ldots, \lambda_U$ of the correlation matrix, which satisfy
$\sum_{u=1}^{U}\lambda_u = 1$ from Eq.~\eqref{C normalization}, the SVD entropy $S$ is
defined as
\begin{equation}
	\label{EQUATION-classical-svdentropy}
	S = -\sum_{u=1}^U \lambda_u \log \lambda_u.
\end{equation}
Computation of $S$ in any classical means requires the diagonalization of the $N_s\times N_s$
matrix $C$, hence its computational complexity is $O(N_s^3)$.

The SVD entropy $S$ has been proposed as an indicator to detect financial 
crisises, such as financial crashes and bubbles, based on the methodology 
of information theory \cite{caraiani2014predictive}.
In the information theory, entropy measures the randomness of random variables \cite{cover1999elements}.
According to the Efficient Market Hypothesis \cite{samuelson1965proof,fama1970efficient}, financial markets during normal periods show highly random behavior, which lead to large entropy.
In fact, it is known that the eigenvalue distribution of the correlation matrix $C$ can be well explained by the chiral random matrix theory, except for some large eigenvalues~\cite{laloux1999noise,plerou1999universal,plerou2002random,utsugi2004random}.
The eigenvector associated with the largest eigenvalue is interpreted as the market portfolio,
which consists of the entire stocks, and the eigenvectors associated with the other large eigenvalues
are interpreted as the portfolios of stocks belonging to different industrial sectors.
The structural relation of these eigenvalues does not change a lot during the normal periods, but it
changes drastically at a point related to financial crisis.
Actually, the stock prices across the entire market or the clusters of several industrial sectors have been reported to show collective behavior when financial crisis occured \cite{onnela2003dynamic,risso2008informational,kenett2011index,nobi2014correlation,ren2014dynamic,zhao2016structure,yin2017trend};
mathematically, this means that the eigenvalues become nearly degenerate, or roughly speaking
the probability distribution of the eigenvalues visibly becomes sharp.
As a result, the SVD entropy $S$ takes a relatively small value, indicating the financial crisis.
This behavior is analogous to that of the statistical mechanical systems near a critical point, where
spins form a set of clusters due to the very large correlation length~\cite{dutta2015quantum}.
Lastly we add a remark that a similar and detailed analysis for image processing can be found in
Ref.~\cite{matsueda2011renormalization}.

\subsection{Computation on quantum devices}
\label{SUBSECTION-quantum-svdentropy}

Computation of the SVD entropy on a quantum device can be performed by firstly training the
PQC $U(\boldsymbol{\theta})$ by the AAE algorithm, to generate a target state in which the stock data
$a_{jt}$ is suitably embedded.
Next, the PQC $U_{\rm SVD}(\xi)$ is variationally trained so that it performs the SVD.
Finally, the SVD entropy is estimated from the output state of the entire circuit
$U_{\rm SVD}(\xi)U(\boldsymbol{\theta})$. In the following, we show the detail discussions of the
procedures.

\subsubsection{The first part: data loading}

The AAE serves as the first part of the entire algorithm; that is, it is used to
load the normalized logarithmic rate of return of the stock price data $a_{jt}$ given in Eq.~\eqref{EQUATION-stock-coefficient}
into a quantum state.
The target state that the AAE aims to approximate is the following bipartite state:
\begin{equation}
	\label{EQUATION-covariance-state}
	\data = \sum_{j=1}^{N_s}\sum_{t=1}^{T}a_{jt}|j \rangle_{\stockdim}|t\rangle_{\timedim},
\end{equation}
where $\{ |j\rangle_{\stockdim} \}_{j=1, \ldots, N_s}$ and $\{ |t\rangle_{\timedim} \}_{t=1, \ldots, T}$
are the computational basis set constructing the stock index Hilbert space
${\cal H}_{\rm stock}$ and the time index Hilbert space ${\cal H}_{\rm time}$,
respectively.
The number of qubits needed to prepare this state is $n_s + n_t$, where
$n_s=O(\log N_s)$ and $n_t=O(\log T)$, meaning that the quantum approach has
an exponential advantage in the memory resource.
Also note that the state \eqref{EQUATION-covariance-state} is normalized
because of Eq.~\eqref{C normalization}.
The partial trace over ${\cal H}_{\rm time}$ gives rise to
\begin{equation}
	\label{EQUATION-partial-trace-psi}
	\rho_{\rm stock}
	= \mathrm{Tr}_{{\cal H}_{\rm time}}(\data\dataket)
	= \sum_{jk}C_{jk}|j\rangle_{\stockdim}\langle k|_{\stockdim},
\end{equation}
where $C_{jk}$ is the $(j,k)$ element of the correlation matrix given in
Eq.~\eqref{EQUATION-covariance-matrix}; that is, $\rho_{\rm stock}=C$ is
realized on a quantum device.
Hence, we need an efficient algorithm to diagonalize $\rho_{\rm stock}$ and
eventually compute the SVD entropy $S$ on a quantum device, and this is the
reason why we use the qSVD algorithm.

\subsubsection{The second part: Quantum singular value decomposition}

We apply the qSVD algorithm \cite{bravo2020quantum} to achieve the above-mentioned
diagonalization task.
The point of this algorithm lies in the  fact that diagonalizing
$\rho_{\rm stock}=C$ is equivalent to realizing the Schmidt decomposition
of $\data$:
\begin{equation}
	\label{schmidt of data}
	\data = \sum_{m=1}^{M} c_m |v_m\rangle_{\stockdim} |v'_m\rangle_{\timedim},
\end{equation}
where $\{c_m \}_{m=1}^M$ are the Schmidt coefficients, and $\{|v_m\rangle_{\stockdim}\}_{m=1}^M$ and $\{|v'_m\rangle_{\timedim}\}_{m=1}^M$ are set of orthogonal
states, with $M \leq{\rm min}(N_s, T)$; note that in general these are not
the computational basis.
Actually, in this representation, $\rho_{\rm stock}$ is calculated as
\begin{equation}
\begin{split}
\label{EQUATION-diagonal}
	\rho_{\rm stock}
	&= \mathrm{Tr}_{{\cal H}_{\rm time}}(\data\dataket) \\
	&= \sum_{m=1}^M |c_m|^2 |v_m\rangle_{\stockdim} \langle v_m|_{\stockdim},
\end{split}
\end{equation}
which is exactly the diagonalization of $\rho_{\rm stock}=C$. 
This equation tells us that the eigenvalue of the correlation matrix 
$C$ is now found to be $\lambda_j=|c_j|^2$ for all $j=1, \ldots, M=U$, 
and thus we end up with the expression
\begin{equation}
	\label{EQUATION-quantum-svd-entropy}
	S = -\sum_{m=1}^{M} |c_m|^2 \log |c_m|^2.
\end{equation}
This coincides with the entanglement entropy between ${\cal H}_{\rm stock}$ 
and ${\cal H}_{\rm time}$; i.e., von Neumann entropy of $\rho_{\rm stock}$, 
$S=-{\rm Tr}(\rho_{\rm stock} \log \rho_{\rm stock})$.

Note now that we cannot efficiently extract the values of $|c_m|^2$ 
from the state $|Data\rangle$, because $\{|v_m\rangle\}_{m=1}^M$ and $\{|v_m^{\prime}\rangle\}_{m=1}^M$ 
are not the computational basis. 
Thus, as the next step, we need to transform the basis $\{|v_m\rangle\}_{m=1}^M$ and 
$\{|v_m^{\prime}\rangle\}_{m=1}^M$ to the computational basis, which is done by using 
qSVD \cite{bravo2020quantum}.

The qSVD is a variational algorithm for finding the transformation that transforms Schmidt basis to the computational basis. For simplicity, let us assume $n_{\rm s} = n_{t}$, which is the case in our numerical demonstration in Section~\ref{SECTION-demonstration}.
Let $|\widetilde{Data}\rangle$ be the output of the AAE circuit, which approximates the target
$\data$.
We train PQCs $U_1(\bold{\xi})$ and $U_2(\bold{\xi^{\prime}})$ with
parameters $\xi$ and $\xi'$, so that, ideally, they realize
\begin{equation}
	\label{EQUATION-svd-state}
	U_1(\xi) \otimes U_2(\xi^{\prime}) |\widetilde{Data}\rangle
	= \sum_{m=1}^{M} c_m|\bar{m}\rangle_{\stockdim} |\bar{m}\rangle_{\timedim}.
\end{equation}
Here, $\{|\bar{m}\rangle_{\stockdim}\}_{m=1}^M$ and $\{|\bar{m}\rangle_{\timedim}\}_{m=1}^M$ are {\it subset} of the {\it computational basis} states, which thus
satisfy $\langle \bar{m} |\bar{\ell}\rangle = \delta_{m, \ell}$ (here we omit the subscript `$\stockdim$' or `$\timedim$' for simplicity).
Clearly, then, the Schmidt basis is identified as
$|v_m\rangle_{\stockdim} = U_1^\dagger(\bold{\xi})|\bar{m}\rangle_{\stockdim}$ and
$|v'_m\rangle_{\timedim} = U_2^\dagger(\bold{\xi'})|\bar{m}\rangle_{\timedim}$.
The training policy is chosen so that $U_1(\xi) \otimes U_2(\xi^{\prime}) |\widetilde{Data}\rangle$
is as close to the Schmidt form in the computational basis as possible.
The cost function to be minimized, proposed in \cite{bravo2020quantum}, is the sum of Hamming
distances between the stock bit sequence and the time bit sequence, obtained as the result of
computational-basis measurement on ${\cal H}_{\rm stock}$ and ${\cal H}_{\rm time}$;
actually, if we measure the right hand side of Eq.~\eqref{EQUATION-svd-state}, the outcomes are
perfectly correlated, e.g., 010 on ${\cal H}_{\rm stock}$ and 010 on ${\cal H}_{\rm time}$.
The cost function is
represented as
\begin{equation}
\label{EQUATION-qsvd-cost}
	{\cal L}_{\rm SVD}(\xi, \xi')
	= \sum_{q=1}^{n_{s}} \frac{1 - \langle \sigma_z^{q} \sigma_z^{q+n_{s}}\rangle}{2},
\end{equation}
where the expectation $\langle \cdot \rangle$ is taken over
$U_1(\xi) \otimes U_2(\xi^{\prime}) |\widetilde{Data}\rangle$.
The operator $\sigma_z^q$ is the Pauli $Z$ operator that acts on the $q$-th qubit.
We see that ${\cal L}_{\rm SVD}(\xi, \xi')=0$ holds, if and only if
$U_1(\xi) \otimes U_2(\xi^{\prime}) |\widetilde{Data}\rangle$ takes the form of the right hand side
of Eq.~\eqref{EQUATION-svd-state}.
Therefore, by training $U_1(\xi)$ and $U_2(\xi^{\prime})$ so that ${\cal L}_{\rm SVD}(\xi, \xi')$ is minimized, we obtain the state that best approximates the Schmidt decomposed state.

Lastly, we gain the information on the amplitude of the output of qSVD
circuit (i.e., the values approximating $|c_m|^2$), via the computational
basis measurements, and then compute the SVD entropy $S$.
For example, we take the method proposed in \cite{li2018quantum}, which
effectively estimates $S$ from the state $\sum_{m=1}^{M} c_m|\bar{m}\rangle_{\stockdim}|\bar{m}\rangle_{\timedim}$;
more specifically, this algorithm utilizes the amplitude estimation
\cite{brassard2002quantum} to estimate $S$ with complexity $\tilde{O}\left(\sqrt{{\rm min}(N_s,T)}/\epsilon^2\right)$,
where $\epsilon$ is the estimation error and the $\tilde{O}$ hides the \mbox{poly}log factor.
The computational complexity of estimating $S$ is negligible if the quantum
state after the qSVD is sparse, or when $T$ is small.
This is indeed the case in our problem for computing the SVD entropy in the
financial example, because in practice only a few large eigenvalues are
associated with the market sectors and carry important information,
especially in an abnormal period  \cite{laloux1999noise,plerou1999universal,plerou2002random,utsugi2004random}.
This situation is well suited to the spirit of the qSVD algorithm, which aims
to estimate only large eigenvalues.
Hence, taking those fact into consideration, we may be able to reduce the
complexity for estimating the value of the SVD entropy.


\subsection{Complexity of the algorithm}

The complexity for computing the SVD entropy can be obtained by setting
$N=N_s T$, $N_{alg} = O(\mbox{poly}(\log N_s T))$, and
$N_{mes} = \tilde{O}\left(\sqrt{{\rm min}(N_s,T)}/\epsilon^2\right)$
in Table~\ref{TABLE-complexity}, where the data is dense.
As noted in Section~\ref{SECTION-algorithm}-C, even though we need
$O(N_s T\log N_s T)$ computation in a classical computing device, the exact data
loading method also requires the same amount of classical computation for compiling
the data into gate operations.
On the other hand, for the execution stage, AAE requires
$O\left(\mbox{poly}(\log(N_s T))\right) \cdot \tilde{O}\left( \sqrt{{\rm min}(N_s,T)}/\epsilon^2\right)$ gate operations in a quantum computing device.
In contrast, for exactly encoding the data we need $O(N_s T)$ gates (say, with the
technique in~\cite{plesch2011quantum,Bergholm_2005}), and the total gate operations
on a quantum computing device is
$O\left(N_s T\right) \cdot \tilde{O}\left( \sqrt{{\rm min}(N_s,T)}/\epsilon^2\right)$,
which is much larger than the one using the AAE method.


\subsection{Demonstration}
\label{SECTION-demonstration}

\begin{table*}[ht]
	\caption{Stock prices for Exxon Mobil Corporation (XOM), Walmart (WMT), Procter \& Gamble (PG), and Microsoft (MSFT) between April 2008 and March 2009.
	}
	\scalebox{1.1}{
		\begin{tabular}{|c|c|c|c|c|c|c|c|c|c|c|c|c|}
			\hline
			Symbol & Apr 08 & May 08 & Jun 08 & Jul 08 & Aug 08 & Sep 08 & Oct 08 & Nov 08 & Dec 08 & Jan 09 & Feb 09 & Mar 09 \\ \hline
			XOM    & 84.80  & 90.10  & 88.09  & 87.87  & 80.55  & 78.04  & 77.19  & 73.45  & 77.89  & 80.06  & 76.06  & 67.00  \\ \hline
			WMT    & 53.19  & 58.20  & 57.41  & 56.00  & 58.75  & 59.90  & 59.51  & 56.76  & 55.37  & 55.98  & 46.57  & 48.81  \\ \hline
			PG     & 70.41  & 67.03  & 65.92  & 60.55  & 65.73  & 70.35  & 69.34  & 64.72  & 63.73  & 61.69  & 54.00  & 47.32  \\ \hline
			MSFT   & 28.83  & 28.50  & 28.24  & 27.27  & 25.92  & 27.67  & 26.38  & 22.48  & 19.88  & 19.53  & 17.03  & 15.96  \\ \hline
		\end{tabular}
	}
	\label{TABLE-stock-data}
\end{table*}

\fig{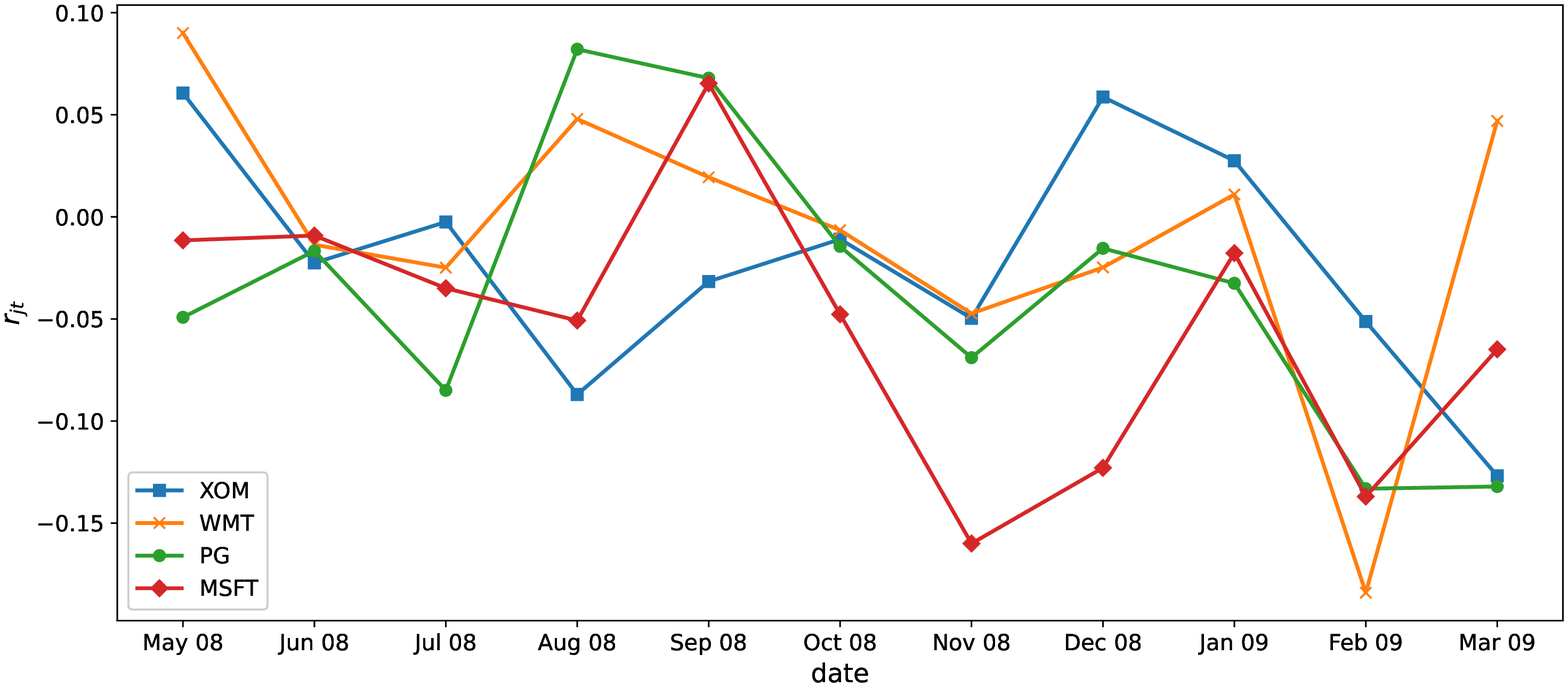}
{Logarithmic rate of return ($r_{jt}$) for each stock and each moment that is computed
	with the data in TABLE \ref{TABLE-stock-data}.
}{FIGURE-stock-data}{width=510pt}

\fig{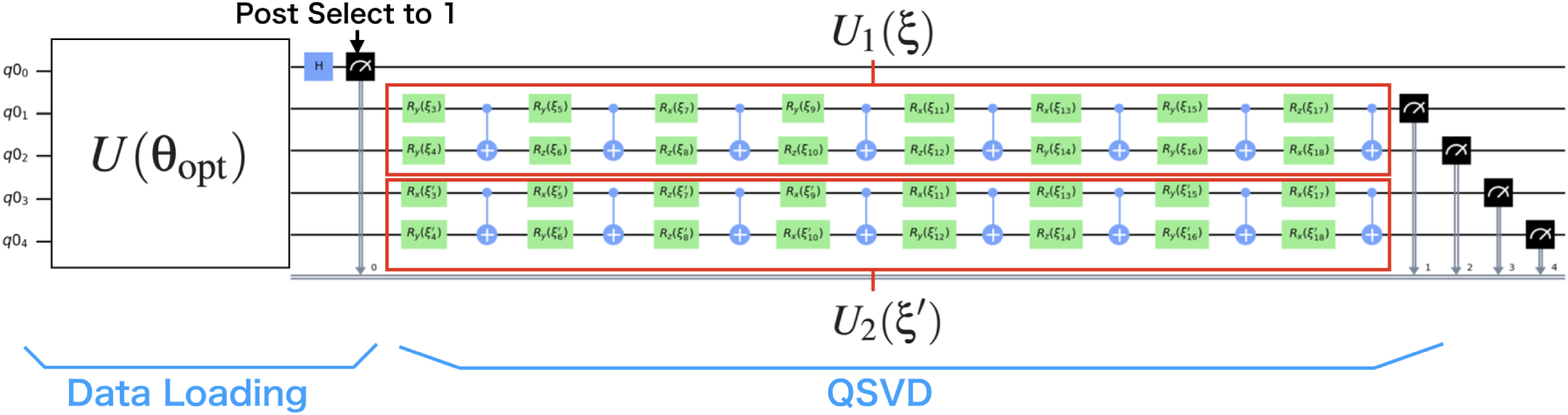}{
Structure of the qSVD circuit.
The parameters $\theta_{\rm opt}$ in the AAE circuit are fixed.
Each layer of $U_1$ is composed of parameterized single-qubit rotational gates
$\exp(-i \xi_r \sigma_{a_r} /2)$ and CNOT gates that connect adjacent qubits, where $\xi_r$ is
the $r$-th parameter and $\sigma_{a_r}$ is the Pauli operator ($a_r = x, y, z$).
The circuit $U_2$ has the same structure as $U_1$. For each trial, we randomly initialize the gate types and parameters, e.g., as for $U_1$, we choose the gate types $\sigma_{a_r}$ ($a_r = x, y, z$) at the beginning of each trial and fix them during training, and we initialize parameters $\xi_r$. We initialize $U_2$ in the same way.}{FIGURE-svd-circuit}{width=520pt}

\fig{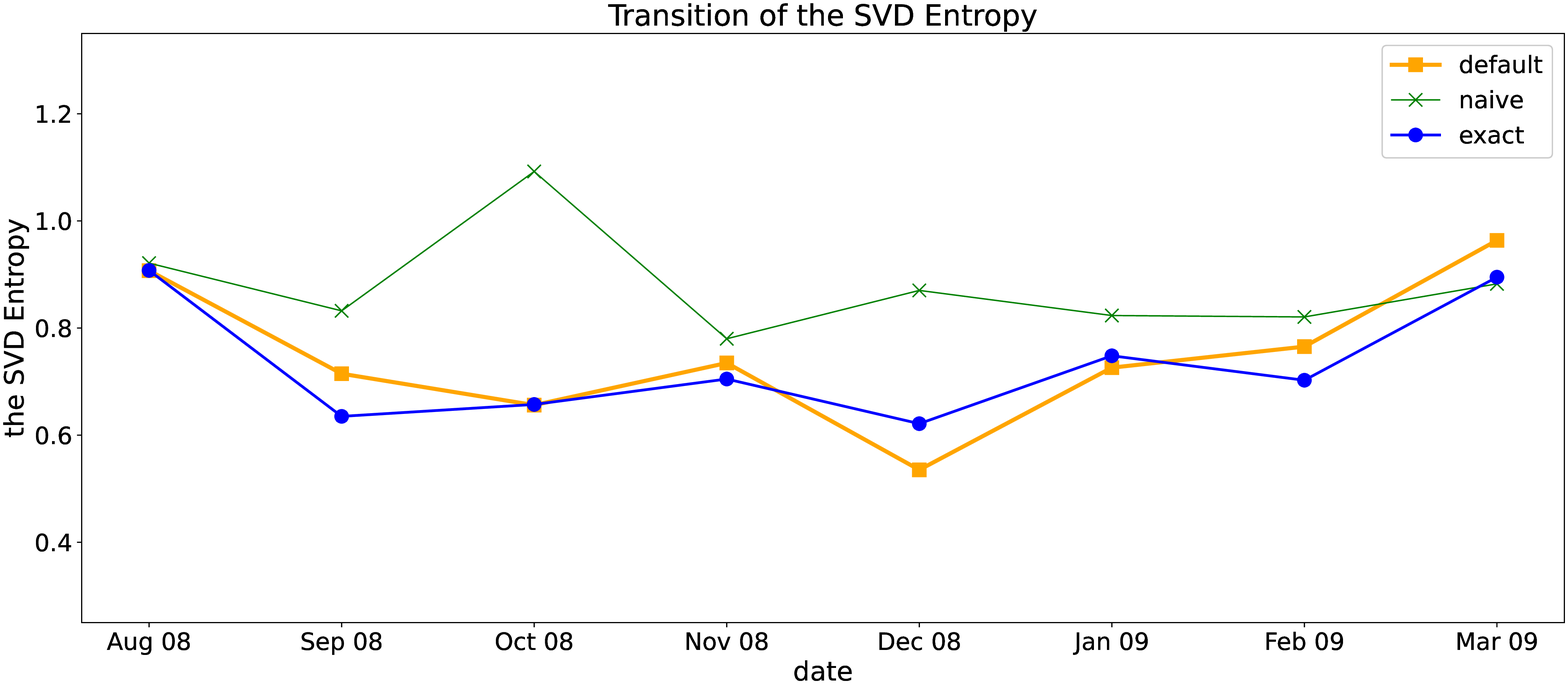}{
Change of the SVD entropy for each term, with different computing method.
The SVD entropy computed via AAE and qSVD algorithms, is shown by the orange
line with square dots.
The exact value of SVD entropy, computed by diagonalizing the correlation
matrix, is shown by the blue line with circle dots.
The SVD entropy computed with the method \cite{zoufal2019quantum} is shown
by the green line with cross marks.}
{FIGURE-svd-entropy}
{width=500pt}

Here we give a numerical demonstration to show the performance of our algorithm composed of
AAE and qSVD, in the problem of computing the SVD entropy for the following stock
data found in the Dow Jones Industrial Average at the end of 2008;
Exxon Mobil Corporation (XOM), Walmart (WMT), Procter \& Gamble (PG), and Microsoft (MSFT). They are top 4 stocks included in Dow Jones Industrial Average by market capitalization at the end of 2008.
For each stock, we use the one-year monthly data from April 2008 to March 2009, which is shown in 
TABLE \ref{TABLE-stock-data}. Data was taken from Yahoo Finance (in every month, the opening price
is used). Fig.~\ref{FIGURE-stock-data} shows the logarithmic rate of return (\ref{EQUATION-logarithm}) for
each stock at every month computed with the data in TABLE \ref{TABLE-stock-data}.

The goal is to compute the SVD entropy at each term, with the length $T=5$ months.
For example, we compute the SVD entropy at August 2008, using the data from April 2008 to
August 2008.
The stock indices $j=1, 2, 3, 4$ correspond to XOM, WMT, PG, and MSFT, respectively.
Also, the time indices $t = 0, 1, 2, 3, 4$ identify the month in which the SVD entropy is computed;
for instance, the SVD entropy on August 2008 is computed, using the data of April 2008 ($t=0$),
May 2008 ($t=1$), June 2008 ($t=2$), July 2008 ($t=3$), and August 2008 ($t=4$).
As a result, $s_{jt}$ has totally $20 = 4~({\rm stocks}) \times 5~({\rm terms})$ components, and thus,
from Eq.~\eqref{EQUATION-logarithm}, both $r_{jt}$ and $a_{jt}$ have $16 = 4 \times 4$ components,
where the indices run over $j=1, 2, 3, 4$ and $t = 1, 2, 3, 4$;
that is, ${\cal H}_{\rm stock}\otimes{\cal H}_{\rm time}={\bf C}^4\otimes{\bf C}^4$.
Note that $\{a_{jt}\}$ contain both positive and negative quantities, and thus AAE algorithm
for Case 2 is used for the data loading.
Hence, we need an additional ancilla qubit, meaning that the total number of qubit is 5.
The extended target state \eqref{EQUATION-sign-manipulated-state} is now given by
\begin{equation}
	|\Bar{\psi}\rangle  =  \sum_{k=0}^{31} \bar{\psi}_k |k\rangle,
\end{equation}
where
\begin{equation*}
	\bar{\psi}_k = \left\{\begin{array}{ccll}
		 & a_{jt}  & {\rm if}\ k=8(j-1) + 2(t-1),\      & a_{jt} \geq 0
		\\
		 & 0       & {\rm if}\ k=8(j-1) + 2(t-1),\      & a_{jt} < 0
		\\
		 & -a_{jt} & {\rm if}\ k=8(j-1) + 2(t-1) + 1,\  & a_{jt} < 0
		\\  & 0 & {\rm if}\ k=8(j-1) + 2(t-1) + 1,\ &a_{jt} \geq 0.
	\end{array}
	\right.
\end{equation*}
The binary representation of $k$ corresponds to the state of the qubits, e.g.,
$|2\rangle \equiv |00010\rangle$.
Then the conditions of perfect data loading, given by Eqs.~\eqref{EQUATION-casetwo-condition-1}
and \eqref{EQUATION-casetwo-condition-2}, are represented as
\begin{equation}
\label{perfect condition Section Demo}
	\begin{split}
		|\langle k|U(\boldsymbol{\theta})|0\rangle^{\otimes 5}|^2 &= \bar{\psi}_k^2, \\
		|\langle k|H^{\otimes 5}U(\boldsymbol{\theta})|0\rangle^{\otimes 5}|^2 &=
		\left(\sum_{\ell=0}^{31}\bar{\psi}_{\ell}\langle \ell|H^{\otimes 5}|k\rangle\right)^2.
	\end{split}
\end{equation}
The right-hand side of these equations are the target probability distributions
to be approximated by the output probability distributions (left-hand side) of
the trained PQC $U(\boldsymbol{\theta})$; that is,
$q_\theta(k) = |\langle k|U(\boldsymbol{\theta})|0\rangle^{\otimes 5}|^2$,
$q_\theta^H(k) = |\langle k|H^{\otimes 5}U(\boldsymbol{\theta})|0\rangle^{\otimes 5}|^2$, $p(k)=\Bar{\psi}_k^2$, and
$p^H(k)=\left(\sum_{\ell=0}^{31}\Bar{\psi}_{\ell}\langle \ell|H^{\otimes 5}|k\rangle\right)^2$.

In this work, we execute the AAE algorithm as follows.
The PQC is the 8-layers ansatz $U(\boldsymbol{\theta})$ illustrated in Fig.~\ref{FIGURE-data-circuit}.
Each layer is composed of the set of parameterized single-qubit rotational gate
$R_y(\theta_r)=\exp(-i \theta_r \sigma_y /2)$ and CNOT gates that connect adjacent qubits;
$\theta_r$ is the $r$-th parameter and $\sigma_y$ is the Pauli $Y$ operator (hence $U(\boldsymbol{\theta})$
is a real matrix).
We randomly initialize all $\theta_r$ at the beginning of each training.
As the kernel function, $\kappa(x, y) = \exp\left(-(x-y)^2/0.25\right)$ is used.
To compute the $r$-th gradient of $\mathcal{L}$ given in Eq.~\eqref{EQUATION-cost-gradient}, we
generate 400 samples for each $q_{\theta_r}^{+}$, $q_{\theta_r}^{-}$, $q_{\theta_r}^{H+}$, and
$q_{\theta_r}^{H-}$.
As the optimizer, Adam \cite{adam} is used; the learning rate is 0.1 for the first $100$ iterations and
0.01 for the other iterations.
The number of iterations (i.e., the number of the updates of the parameters) is set to $2     00$ for
training $U(\boldsymbol{\theta})$.
We performed 10 trials for training $U(\boldsymbol{\theta})$ and then chose the one which best minimizes the
cost $\mathcal{L}$ at the final iteration step.

\begin{table*}[ht]
\centering
\caption{The mean value and the maximum value of the overlap $\mathcal{O} = \left|\langle 1|\langle Data |VU(\boldsymbol{\theta})|0\rangle^{\otimes 5}\right|$ for
each term, depending on the values of the cost functions $\mathcal{L}_1$
and $\mathcal{L}_2$.
The number of trials that satisfy each condition out of 10 trials is also
listed in the table.}
\begin{tabular}{|l|c|c|c|c|c|c|}
\hline
\multirow{3}{*}{Term} &
  \multicolumn{3}{c|}{\begin{tabular}[c]{@{}c@{}}$\mathcal{L}_1<0.01$ and $\mathcal{L}_2<0.01$\\ at the final iteration\end{tabular}} &
  \multicolumn{3}{c|}{\begin{tabular}[c]{@{}c@{}}otherwise: $\mathcal{L}_1\geq 0.01$ or $\mathcal{L}_2 \geq 0.01$\\ at the final iteration\end{tabular}} \\ \cline{2-7}
 &
  \multirow{2}{*}{\begin{tabular}[c]{@{}c@{}}\# of trials\\ satisfying the condition\end{tabular}} &
  \multicolumn{2}{c|}{$\mathcal{O}$} &
  \multirow{2}{*}{\begin{tabular}[c]{@{}c@{}}\# of trials\\ satisfying the condition\end{tabular}} &
  \multicolumn{2}{c|}{$\mathcal{O}$} \\ \cline{3-4} \cline{6-7}
                &   &  mean  & max   &   &  mean   & max   \\ \hline
Apr 08 - Aug 08 & 2 & 0.977 & 0.981 & 8 & 0.591 & 0.851 \\ \hline
May 08 - Sep 08 & 4 & 0.948 & 0.973 & 6 & 0.435 & 0.646 \\ \hline
Jun 08 - Oct 08 & 4 & 0.960 & 0.977 & 6 & 0.284 & 0.718 \\ \hline
Jul 08 - Nov 08 & 3 & 0.973 & 0.981 & 7 & 0.439 & 0.875 \\ \hline
Aug 08 - Dec 08 & 2 & 0.968 & 0.972 & 8 & 0.437 & 0.648 \\ \hline
Sep 08 - Jan 08 & 3 & 0.955 & 0.968 & 7 & 0.203 & 0.441 \\ \hline
Oct 08 - Feb 08 & 7 & 0.957 & 0.980 & 3 & 0.578 & 0.829 \\ \hline
Nov 08 - Mar 08 & 7 & 0.969 & 0.979 & 3 & 0.613 & 0.655 \\ \hline
\end{tabular}
\label{TABLE-overlap}
\end{table*}

Suppose that the above AAE algorithm generated the quantum state $|\widetilde{Data}\rangle$,
which approximates Eq.~\eqref{EQUATION-covariance-state}.
Then the next step is to apply the qSVD circuit to $|\widetilde{Data}\rangle$ and then compute
the SVD entropy.
The PQCs $U_1(\bold{\xi})$ and $U_2(\xi^{\prime})$, which respectively act on the stock
state $|j\rangle$ and the time state $|t\rangle$, are set to 2-qubits 8-layers ansatz
illustrated in Fig~\ref{FIGURE-svd-circuit}.
Each layer of $U_1$ is composed of parameterized single-qubit rotational gates
$\exp(-i \xi_r \sigma_{a_r} /2)$ and CNOT gates that connect adjacent qubits, where $\xi_r$ is
the $r$-th parameter and $\sigma_{a_r}$ is the Pauli operator ($a_r = x, y, z$).
As seen in the figure, $U_2$ has the same structure as $U_1$. For each trial, we randomly initialize the gate types and parameters, e.g., as for $U_1$, we choose the gate types $\sigma_{a_r}$ ($a_r = x, y, z$) at the beginning of each trial and fix them during the training, and we initialize parameters $\xi_r$. We initialize $U_2$ in the same way.
Also we used Adam optimizer, with learning rate 0.01.
For simulating the quantum circuit, we used Qiskit \cite{qiskit}.
To focus on the net approximation error that stems from qSVD algorithm, we assume that
the gradient of ${\cal L}_{\rm SVD}$ can be exactly computed (equivalently, infinite number of
measurements are performed to compute this quantity).
The number of iterations for training $U_1(\xi) \otimes U_2(\xi^{\prime}) \data$ is $500$.
Unlike the case of AAE, we performed qSVD only once, to determine the optimal parameter set
$(\xi_{\rm opt}, \xi_{\rm opt}')$.
Finally, we compute the SVD entropy based on the amplitude of the final state
$U_1(\xi_{\rm opt}) \otimes U_2(\xi_{\rm opt}^{\prime}) |\widetilde{Data}\rangle$, under
the assumption that the ideal quantum state tomography can be executed.

\fig{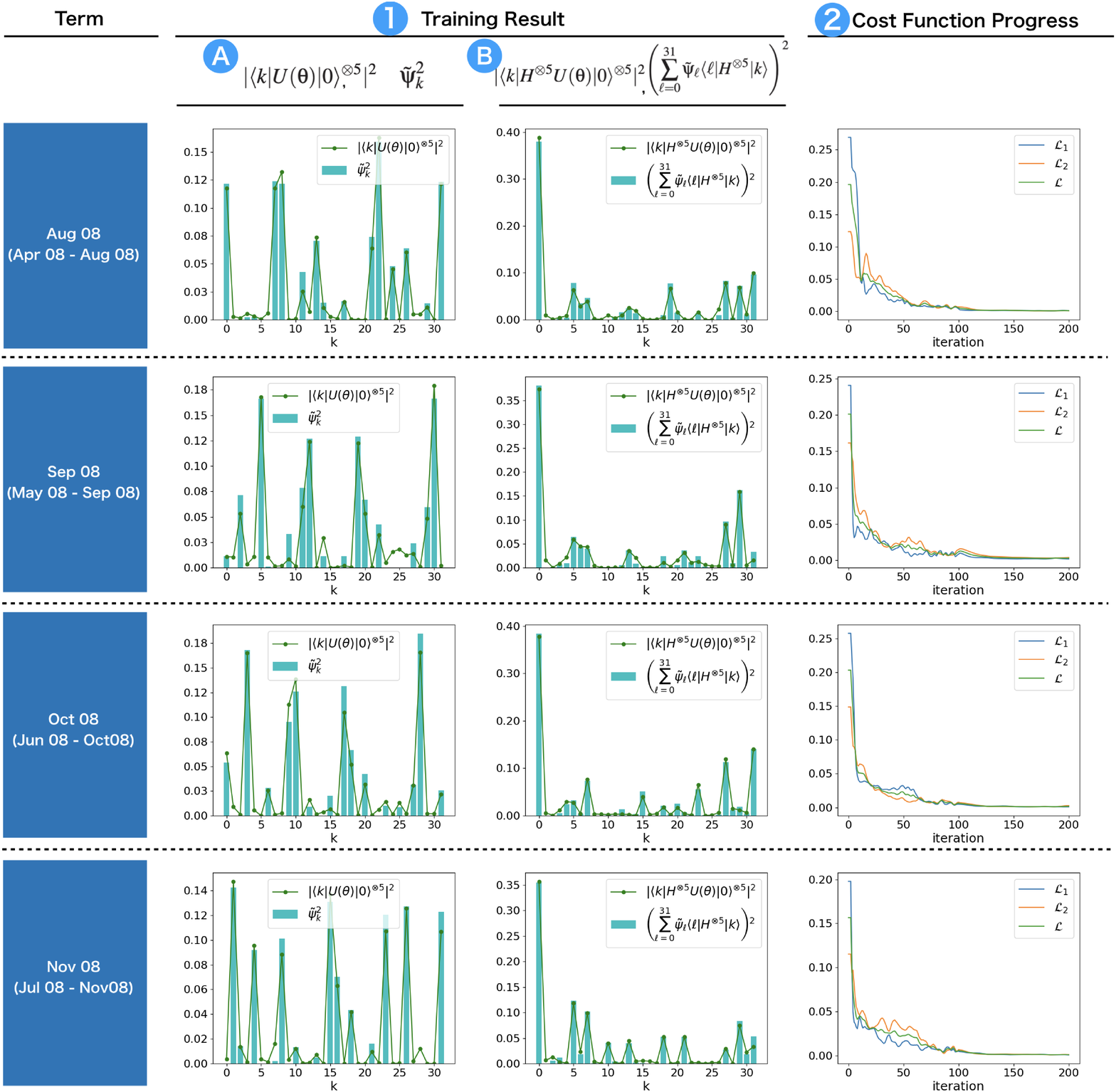}
{
\textcircled{\scriptsize 1}:Example of the training results of
$q_\theta(k)=|\langle k|U(\boldsymbol{\theta})|0\rangle^{\otimes 5}|^2$ and
$q^H_\theta(k)=|\langle k|H^{\otimes 5}U(\boldsymbol{\theta})|0\rangle^{\otimes 5}|^2$
for each term (green lines), and the corresponding target distributions  $p(k)=\Tilde{\psi}_k^2$ and
$p^H(k)=\left(\sum_{\ell=0}^{31}\tilde{\psi}_{\ell}\langle \ell|H^{\otimes 5}|k\rangle\right)^2$ (green bars).
These distributions are the best one in the sense that the cost function
$\mathcal{L}$ at the 200-th epoch takes the smallest in 10 trials. \textcircled{\scriptsize 2}: The change of the cost function $\mathcal{L}$
in the same trial as \textcircled{\scriptsize 1} for each term.
In the same figures, we also show the change of the cost functions for the
two distributions $q_\theta(k)$ and $q^H_\theta(k)$ that contribute to $\mathcal{L}$.}
{FIGURE-training-quality}{width=540pt}

The SVD entropy in each term, computed through AAE and qSVD algorithms, is shown
by the orange line with square dots in Fig.~\ref{FIGURE-svd-entropy}.
As a reference, the exact value of SVD entropy, computed by diagonalizing the correlation matrix,
is shown by the blue line with circle dots.
Also, to see a distinguishing property of AAE, we study the naive data-loading method \cite{zoufal2019quantum} that trains the PQC $U_{\rm naive}(\boldsymbol{\theta})$ so that it learns only the absolute value of the data by minimizing the
cost function
$\mathcal{L}_{\rm MMD}(|\langle j|\langle  t|U_{\rm naive}(\boldsymbol{\theta})|0\rangle^{\otimes 4}|^2, a_{jt}^2)$;
namely, $U_{\rm naive}(\boldsymbol{\theta})$ loads the data so that the absolute values of the amplitudes of
$U_{\rm naive}(\boldsymbol{\theta})|0\rangle^{\otimes 4}$ is close to $|a_{jt}|$ yet without taking into account the signs.
The resulting value of SVD entropy computed with this naive method is shown by the green line with
cross marks.
Importantly, the SVD entropies computed with our AAE algorithm well approximate the exact
values, while the naive method poorly works at some point of term.
Note that the estimation errors in the case of AAE is within the acceptable range for application,
because the SVD entropy usually fluctuates by several percent during the normal period, while it
can change drastically by a few tens of percent at around financial
events \cite{caraiani2014predictive,gu2015does,civitarese2016volatility,caraiani2018modeling}.


Now, to see the quality of the data loading circuit for each trial in detail, we
compute the overlap between the target state $\data$ and the generated state $VU(\boldsymbol{\theta})|0\rangle^{\otimes 5}$ after each training (10 trials for each term). The overlap can be measured by using the following value
\begin{equation}
    \mathcal{O} \equiv \left|\langle 1|\langle Data |VU(\boldsymbol{\theta})|0\rangle^{\otimes 5}\right|
\end{equation}
at the final iteration of each trial.
In fact, in terms of $\mathcal{O}$, the generated state can be represented as
\begin{equation}
    V U(\boldsymbol{\theta})|0\rangle = \Big( \mathcal{O}\data + \sqrt{1 - \mathcal{O}^2}|Data^{\perp}\rangle \Big) |1\rangle,
\end{equation}
where $|Data^{\perp}\rangle$ is a state that is orthogonal to $\data$.
Namely, the closer the value of $\mathcal{O}$ is to 1, the more accurately $VU(\boldsymbol{\theta})$ generates $\data$.
To evaluate the statistics of the overlap in each trial, we divide the 10 trials for each term into the following two patterns of conditions satisfied by the
cost function at the final iteration step:
$\mathcal{L}_1, \mathcal{L}_2 < 0.01$ or otherwise, where we simply denote
\begin{equation*}
    \mathcal{L}_1 = \mathcal{L}_{MMD}(q_\theta, p), ~~
        \mathcal{L}_2 = \mathcal{L}_{MMD}(q_\theta^H, p^H).
\end{equation*}
Recall that $(q_\theta, p, q_\theta^H, p^H)$ are given below
Eq.~\eqref{perfect condition Section Demo}.
In Table~\ref{TABLE-overlap}, we show the mean value and the maximum value of
the overlap $\mathcal{O}$ for each pattern and for each term.
The number of trials that satisfy each condition out of 10 trials is also listed
in the same table.
We then find that, as long as the condition $\mathcal{L}_1, \mathcal{L}_2 < 0.01$ is satisfied, the mean of $\mathcal{O}$ is larger than $0.94$, and there are
at least 2 out of 10 trials that satisfy this condition.
Also, the maximum value of $\mathcal{O}$ is larger than $0.96$ in all terms;
such large overlaps between the target state and the generated state will lead
to a successful computation of the SVD entropy for each term.
When $\mathcal{L}_1 \geq 0.01$ or $\mathcal{L}_2 \geq 0.01$, on the other hand,
$\mathcal{O}$ takes a relatively small value; in this case the subsequent qSVD
algorithm may yield an imprecise value of SVD entropy, hence this trial should
be discarded.
A notable point here is that the success probability is relatively high;
a thorough examination for a larger system is an important future work.

In Fig.~\ref{FIGURE-training-quality}, we show an example of set of the
training results of $q_\theta(k)$ and $q_\theta^H(k)$ for four terms
Apr\ 08-Aug\ 08, May\ 08-Sep\ 08, Jun\ 08-Oct\ 08, and Jul\ 08-Nov\ 08
(green lines);
this set of distributions is the best one out of 10 trials in the sense
that it minimizes the cost $\mathcal{L}_{\rm MMD}(\boldsymbol{\theta})$ at the 200-th
iteration step, corresponding to the case when the overlap is maximized for
each term.
Also the target distributions $p(k)$ and $p^H(k)$ are illustrated with green
bars.
The right column of Fig.~\ref{FIGURE-training-quality} plots the change of
the costs
$\mathcal{L}_1$, $\mathcal{L}_{2}$, and
$\mathcal{L}
	=(\mathcal{L}_1 + \mathcal{L}_2)/2$ for each term.
These results confirm that the AAE algorithm realizes near perfect data-loading;
that is, the resulting model distributions $q_\theta$ and $q_\theta^H$ well approximate the target distributions $p$ and $p^H$, respectively, which
eventually leads to the successful computation of SVD entropy as discussed
above.

Here we point out the interesting feature of the SVD entropy, which can be observed from Figs.~\ref{FIGURE-stock-data} and \ref{FIGURE-svd-entropy}.
Figure~\ref{FIGURE-stock-data} shows that, until August 2008, the stocks did not strongly correlate
with each other, which leads to the relatively large value of SVD entropy ($\sim 0.9$) as seen in Fig.~\ref{FIGURE-svd-entropy}.
On September 2008, the Lehman Brothers bankruptcy ignited the global financial crisis.
As a result, from October 2008 to February 2009, the stocks became strongly correlated with each
other.
In such a case, many of stocks cooperatively moved as seen in Fig.~\ref{FIGURE-stock-data}.
This strong correlation led to the small SVD entropy ($\sim 0.7$) from October to February,
which is an evidence of the financial crisis.
On March 2009, Fig.~\ref{FIGURE-svd-entropy} shows that the SVD entropy again takes relatively
large value ($\sim 0.9$), indicating that the market returned to normal and each stock moved differently.
Interestingly, according to the S\&P index, it is argued that the financial crisis ended on March 2009
(e.g., see~\cite{Manda10stockmarket}), which is consistent to the result of SVD entropy.
We would like to emphasize that AAE algorithm correctly computed the SVD entropy and enables
us to capture the above-mentioned financial trends.

Lastly, it is surely important to assess the performance of AAE for 
other example problems with different size and data-set. 
Appendix \ref{section:appendix-experiment} gives such a demonstration 
where the number of stocks is eight.

\section{Conclusions}
\label{SECTION-conclusion}

This paper provides the Approximate Amplitude Encoding (AAE) algorithm that effectively loads
a given classical data into a shallow parameterized quantum circuit.
The point of the AAE algorithm is in the formulation of a valid cost function
composed of two types of maximum mean discrepancy measures, based on the perfect
encoding condition (Theorem 1 for Case 1 and Theorem 2 for Case 2);
training of the circuit is executed by minimizing this cost function, which
enables encoding the signs of the data components unlike the previous proposal
(that can only load the absolute values).
We also provide an algorithm composed of AAE and the existing quantum singular
value decomposition (qSVD) algorithm, for computing the SVD entropy in the stock
market.
A thorough numerical study was performed, showing that the approximation error
of AAE was found to be sufficiently small in this case and, as a result, the
subsequent qSVD algorithm yields a good approximation solution.

To show that the proposed AAE algorithm will be practically useful to
implement various quantum algorithms that need classical data loading,
it is important to examine a larger system, e.g., a 20-qubits problem
with 20 layers ansatz.
In fact in this case the number of parameters is 400, while the degree of
freedom of the state vector is $2^{20} \approx 1,000,000$, reflecting that
the polynomial-size circuit could deal with an exponential-size problem.
However, even in this potentially classically-doable size setting, there are
several practical problems to be resolved.
For instance, we expect that the gradient vanishing issue will arise, which
needs careful application of several (existing) methods such as circuit
initialization \cite{Grant2019initialization}, special structured ansatz \cite{cerezo2021cost}, and parameter embedding
\cite{volkoff2021large}.
Moreover, recently we find some approaches for approximating a large circuit
with set of small circuits \cite{tang2021cutqc,peng2020simulating}; these methods are worth investigating to
address the scalability of our method.
At the same time, a notable point of the problem of calculating the SVD entropy
is that it does not require a very precise calculation but only a global trend
over a certain time period.
Hence we need to carefully determine the number of layers as well as the iteration
steps of the variational algorithm to have necessary precision; in particular the
former might be further reduced using existing techniques e.g., \cite{grimsley2019adaptive,tang2021qubit,ryabinkin2018qubit,tkachenko2021correlation}.
With these elaboration, furthermore, we are also interested in testing the
algorithm with a real quantum computing device.
Overall, these additional tasks are all important and yet not straightforward,
so we will study this problem as a separate work.

\section*{Acknowledgments}
This work was supported by Grant-in-Aid for JSPS Research Fellow Grant No.~22J01501,
and MEXT Quantum Leap Flagship Program Grant Number JPMXS0118067285 and JPMXS0120319794.

\appendix       
\section{Proof of Theorem 1}
\label{APPENDIX-proof}
\begin{customthm}{1}
	In Case 1, if (\ref{EQUATION-caseone-condition-1}) and (\ref{EQUATION-caseone-condition-2}) are exactly satisfied, $ U(\boldsymbol{\theta})|0\rangle = \sum_j \dvector_j |j\rangle$ or $ U(\boldsymbol{\theta})|0\rangle = -\sum_j \dvector_j |j\rangle$.
\end{customthm}
\begin{proof}
	Let us denote $\avector_j$ by $\langle j |U(\boldsymbol{\theta})|0\rangle$. Then, (\ref{EQUATION-caseone-condition-1}) and (\ref{EQUATION-caseone-condition-2}) are rewritten as
	\begin{align}
		\label{EQUATION-caseone-condition-3}
		\avector_j^2                                                 & = \dvector_j^2 \qquad(\forall j)                                                 \\
		\label{EQUATION-caseone-condition-4}
		\left(\sum_{k=0}^{N-1} H^{\otimes n}_{jk}\avector_k\right)^2 & = \left(\sum_{k=0}^{N-1} H^{\otimes n}_{jk}\dvector_k\right)^2 \qquad(\forall j)
	\end{align}
	where $H^{\otimes n}_{jk} \equiv \langle j|H^{\otimes n}|k\rangle$. For $j=0$, the left hand side of (\ref{EQUATION-caseone-condition-4}) becomes
	\begin{align}
		\label{EQUATION-theorem1-inequality}
		\left(\sum_{k=0}^{N-1} H^{\otimes n}_{0k}\avector_k\right)^2 = \frac{1}{2^n}
		\left(\sum_{k=0}^{N-1} \avector_k\right)^2
		\leq \frac{1}{2^n}\left(\sum_{k=0}^{N-1} |\avector_k|\right)^2
	\end{align}
	where the equality holds only when $\avector_k \geq 0\ (\forall k)$ or $\avector_k \leq 0\ (\forall k)$. The equality condition is equivalent to $\avector = \dvector$ or $\avector = -\dvector$, because of (\ref{EQUATION-caseone-condition-3}) and the condition of Case 1: $\dvector_j \geq0 \ (\forall j)$ or $\dvector_j \leq0 \ (\forall j)$. Conversely, if the equality condition is not satisfied, we find that
	\begin{equation}
		\begin{split}
		    \frac{1}{2^n}
			\left(\sum_{k=0}^{N-1}
			\avector_k\right)^2 <
			\frac{1}{2^n}
			\left(\sum_{k=0}^{N-1} |\avector_k|\right)^2 &= \frac{1}{2^n} \left(\sum_{k=0}^{N-1} \dvector_k\right)^2 \\ &=\left(\sum_{k=0}^{N-1} H^{\otimes n}_{0k}\dvector_k\right)^2,
		\end{split}
	\end{equation}
	which contradicts to (\ref{EQUATION-caseone-condition-4}) for $j=0$. Thus, the equality condition of (\ref{EQUATION-theorem1-inequality}) is satisfied, i.e., $\avector = \dvector$ or $\avector = -\dvector$.
\end{proof}

\section{Amplification of the success probability in Case 2}
\label{SECTION-amplification}
In Case 2, the encoding can be carried out with success probability
$1/2$ in the ideal case (i.e., the case where Eqs.~(\ref{EQUATION-casetwo-condition-1})
and (\ref{EQUATION-casetwo-condition-2}) are exactly satisfied).
But by applying the amplitude amplification operation \cite{brassard2002quantum},
we obtain $\data$ with success probability $1$, instead of $1/2$, although more
gates to implement this extra operation are required.
The method is described as follows.
By adding another qubit, it holds
\begin{equation}
	\label{EQUATION-grover-target}
	\begin{split}
		&\pm ( I_n\otimes H)U(\boldsymbol{\theta})  |0\rangle^{\otimes n+1}H|0\rangle \\
		&=
		\frac{|Data^{+}\rangle - |Data^{-}\rangle}{2} |00\rangle
		+ \frac{|Data^{+}\rangle - |Data^{-}\rangle}{2}|01\rangle \\
		&+\frac{|Data^{+}\rangle + |Data^{-}\rangle}{2} |10\rangle
		+ \frac{|Data^{+}\rangle + |Data^{-}\rangle}{2}|11\rangle
	\end{split}.
\end{equation}
Similar to \cite{aaronson2020quantum,Nakaji2020FasterAE}, the amplitude
amplification operator $\mathcal{Q}$ can be defined as
\begin{equation*}
	\mathcal{Q} \equiv \mathcal{A}(I_{n + 2}
	- 2|0\rangle_{n+2}\langle0|_{n+2}) \mathcal{A}^{\dag}(I_{n+2}
	- 2I_n\otimes|11\rangle\langle11|),
\end{equation*}
where $\mathcal{A}\equiv ( I_n\otimes H)U(\boldsymbol{\theta})\otimes H$. The operator $\mathcal{Q}$ amplifies the amplitude of the state where the last two qubits are $|11\rangle$. In general, given the amplitude before the amplification as $\cos\xi$, one application of the amplitude amplification operator changes the amplitude to $\cos(3\xi)$. In our case, $\xi=\pi/3$, and therefore the resulting amplitude after the amplification is $\cos(\pi) = -1$.
Thus by applying $\mathcal{Q}$ to the state (\ref{EQUATION-grover-target}), we have
\begin{equation}
	\mathcal{Q}\mathcal{A} |0\rangle^{\otimes n}|0\rangle|0\rangle
	= \pm(|Data^{+}\rangle + |Data^{-}\rangle) |1\rangle|1\rangle = \pm\data |1\rangle|1\rangle.
\end{equation}
Namely, $\data$ is obtained with probability 1 (by ignoring the last two qubits). 

\section{Improvement of AAE}
\label{APPENDIX-study}

In our algorithm we train the data loading circuit by using the measurement results in the computational basis and the Hadamard basis. However, it is possible to use the other basis; namely, given $X$ as an orthogonal operator, we can train $U(\boldsymbol{\theta})$ so that
\begin{align}
	\label{EQUATION-casex-condition-1}
	|\langle j|U(\boldsymbol{\theta})|0\rangle^{\otimes n}|^2  & = \dvector_j^2
	\\
	\label{EQUATION-casex-condition-2}
	|\langle j|XU(\boldsymbol{\theta})|0\rangle^{\otimes n}|^2 & = \left(\sum_{k=0}^{N-1}\dvector_k\langle j|X|k\rangle\right)^2.
\end{align}
Then, the question is whether there exists $X$ such that $U(\boldsymbol{\theta})|0\rangle = \sum_j \dvector_j |j\rangle$ or $U(\boldsymbol{\theta})|0\rangle = -\sum_j \dvector_j |j\rangle$ is satisfied when (\ref{EQUATION-casex-condition-1}) and (\ref{EQUATION-casex-condition-2}) hold. In the following, we show that there exists such $X$ but it is difficult to
find it for arbitrary $\dvector$.

As the preparation, we define the $n\times n$ row switching
matrix $A^{(n)}[j, k]$ as
\scriptsize
\begin{equation}
	A^{(n)}[j,k] =
	\left(\begin{array}{llllllllllll}
			1   &        &   &   &   &        &   &   &   &        &
			\cr & \ddots &   &   &   &        &   &   &   &        &
			\cr &        & 1 &   &   &        &   &   &   &        &
			\cr &        &   & 0 &   &        &   & 1 &   &        &
			\cr &        &   &   & 1 &        &   &   &   &        &
			\cr &        &   &   &   & \ddots &   &   &   &        &
			\cr &        &   &   &   &        & 1 &   &   &        &
			\cr &        &   & 1 &   &        &   & 0 &   &        &
			\cr &        &   &   &   &        &   &   & 1 &        &
			\cr &        &   &   &   &        &   &   &   & \ddots &
			\cr &        &   &   &   &        &   &   &   &        & 1
		\end{array}\right)
	(0 \leq j \neq k \leq n-1)
\end{equation}
\normalsize
that can be created by swapping the row $j$ and the row $k$ of the identity matrix.
Then, given a $n\times n$ matrix $R$, the matrix product $A^{(n)}[j, k]R$ is the matrix produced by exchanging the row $j$ and the row $k$ of $R$ and the matrix product $RA^{(n)}[j, k]$ is the matrix produced by exchanging the column $j$ and the column $k$ of $R$. We denote by $\mathcal{A}(n)$ as the set of the $n\times n$ matrices that can be written as the product of $A^{(n)}[j, k]$s (for example $A^{(8)}[0, 2]A^{(8)}[4, 7]A^{(8)}[2, 3] \in \mathcal{A}(8)$). By using the above notations, we give the following definition for the pair of a vector and a square matrix.
\begin{definition}
	Let $\kvector$ be a column vector, $\mathbf{dim}(\kvector)$ be the number of columns of $\kvector$, and $A$ be a square matrix that has $\mathbf{dim}(\kvector)$ columns/rows.
	The pair ($\kvector$, $A$) is said to be {\bf pair-block-diagonal} if there exists $P, Q \in {\mathcal{A}(\mathbf{dim}(\kvector))}$ that transforms $\kvector$ and $A$ as $\kvector^{\prime} = Q\kvector$ and $A^{\prime} = PAQ$ where $\kvector^{\prime}$ and $A^{\prime}$ are splittable as
	\begin{equation}
		\kvector^{\prime} = \vectorsplit{\kvector} , \qquad A^{\prime} = \matrixsplit{A}
	\end{equation}
	so that $A_{\uparrow\downarrow}\kvector_{\downarrow} = 0, A_{\downarrow\uparrow}\kvector_{\uparrow} = 0$.
	Here $\kvector_{\uparrow}, \kvector_{\downarrow} \neq 0$ and the number of rows in $A_{\uparrow\uparrow}$ and $A_{\uparrow\downarrow}$ is the same as that of $\kvector_{\uparrow}$. The $P$ is said to be a \bold{left-pair-block-generator} and the $Q$ is said to be a \bold{right-pair-block-generator}.
\end{definition}
By using the definition, we can state the following theorem:
\begin{customthm}{3}
	\label{THEOREM-general-amplitude}
	Suppose that X is an $N\times N$ real matrix.
	There exist $N$-element real vectors $\dprime(\neq \dvector, -\dvector)$ and $\cprime = X\dprime$ that satisfy
	\begin{equation}
		\label{EQUATION-amplitude-equality}
		\dprime_j^{2} = \dvector_j^2,
		\cprime_j^{2} = \cvector_j^2
		\qquad (\forall j\in [0,1,\cdots N-1])
	\end{equation}
	if and only if the combination ($\dvector, X$) is pair-block-diagonal where $\cvector$ is an $N$-$element$ vector that satisfies
	\begin{equation}
		\label{EQUATION-relation}
		\cvector = X \dvector.
	\end{equation}
\end{customthm}

\begin{proof}
	Firstly we prove the sufficient condition of the theorem.
	If the sufficient assumption is satisfied, there exist a left-pair-block-generator $P$ and a right-pair-block-generator $Q$. By using $Q^2=I$, (\ref{EQUATION-relation}) can be transformed into
	\begin{equation}
		\label{EQUATION-matrix-pq-form}
		P\cvector = (PXQ) (Q\dvector).
	\end{equation}
	From the definition of pair-block-diagonal, $PXQ$ and $Q\dvector$ are block composed
	\begin{equation}
		\label{EQUATION-matrix-split}
		PXQ = \matrixsplit{X}, \qquad Q\dvector = \vectorsplit{\dvector}
	\end{equation}
	where the number of columns of $X_{\uparrow\uparrow}$ and $X_{\uparrow\downarrow}$ equals to the number of rows of $\dvector_{\uparrow}$, and
	\begin{equation}
		\label{EQUATION-submatrix-condition}
		X_{\uparrow\downarrow}\dvector_{\downarrow} = 0, \qquad X_{\downarrow\uparrow}\dvector_{\uparrow} = 0.
	\end{equation}
	Note that $\dvector_{\uparrow}, \dvector_{\downarrow}\neq 0$ and
	\begin{equation}
		\label{EQUATION-dvector}
		\dvector = Q\left(\begin{array}{l}
				\dvector_{\uparrow} \\
				\dvector_{\downarrow}
			\end{array}
		\right).
	\end{equation}
	Substituting (\ref{EQUATION-matrix-split}) and (\ref{EQUATION-submatrix-condition}) into (\ref{EQUATION-matrix-pq-form}), we get
	\begin{equation}
		P\cvector = \left(\begin{array}{l}
			X_{\uparrow\uparrow}\dvector_{\uparrow} \\
			X_{\downarrow\downarrow}\dvector_{\downarrow}
		\end{array}
		\right),
	\end{equation}
	and therefore,
	\begin{equation}
		\label{EQUATION-cvector}
		\cvector = P\left(\begin{array}{l}
				X_{\uparrow\uparrow}\dvector_{\uparrow} \\
				X_{\downarrow\downarrow}\dvector_{\downarrow}
			\end{array}
		\right)
	\end{equation}
	holds becuase $P^2=I$.
	If we set
	\begin{equation}
		\label{EQUATION-aprime}
		\dprime = Q\left(\begin{array}{cc}
				\dvector_{\uparrow} \\
				-\dvector_{\downarrow}
			\end{array}
		\right)
	\end{equation}
	then
	\begin{equation}
		\label{EQUATION-cprime}
		\cprime = X\dprime = P (P X Q) \left(\begin{array}{cc}
				\dvector_{\uparrow} \\
				-\dvector_{\downarrow}
			\end{array}
		\right) = P\left(\begin{array}{c}
				X_{\uparrow\uparrow}\dvector_{\uparrow} \\
				-X_{\downarrow\downarrow}\dvector_{\downarrow}
			\end{array}
		\right).
	\end{equation}
	Because $P, Q$ are matrices that interchange rows, comparing (\ref{EQUATION-aprime}) and (\ref{EQUATION-dvector}); (\ref{EQUATION-cprime}) and (\ref{EQUATION-cvector}), we see that
	\begin{equation}
		\dprime_j^2 = \dvector_j^2, \qquad
		\cprime_j^2 = \cvector_j^2, \qquad \dprime \neq \dvector, -\dvector
	\end{equation}
	which concludes the proof of the sufficient condition of the theorem.

	Next, we prove the necessity condition of the theorem. If the necessity assumption holds, there exist $\dprime$  and $\cprime$ that satisfy $\cprime = X \dprime$ and (\ref{EQUATION-amplitude-equality}). We set
	\begin{equation}
		\begin{split}
			\cvector^{+} = \frac{\cvector + \cprime}{2},
			\cvector^{-} = \frac{\cvector - \cprime}{2}.
		\end{split}
	\end{equation}
	Then for each element of $\cvector^{-}, \cvector^{+}$, it holds that
	\begin{equation}
		\begin{split}
			\cvector^{+}_j &= \left\{\begin{array}{cl}
				\cvector_j & ({\rm if}\ \cvector_j = \cprime_j)  \\
				0\         & ({\rm if}\ \cvector_j = -\cprime_j)
			\end{array}
			\right. \\
			\cvector^{-}_j &= \left\{\begin{array}{cl}
				0\         & ({\rm if}\ \cvector_j = \cprime_j)  \\
				\cvector_j & ({\rm if}\ \cvector_j = -\cprime_j)
			\end{array}
			\right.
		\end{split}.
	\end{equation}
	We see that $\cvector^{-}_j$ is non-zero only if $\cvector^{+}_j$ is zero, and vice versa. Therefore, by using a row swap matrix $P\in \mathcal{A}(N)$,  $\cvector^{-}_i$ and $\cvector^{+}_i$ can be transformed as
	\begin{equation}
		\label{EQUATION-p-cvector}
		P\cvector^{+} = \left(\begin{array}{cc}
				\cvector^{\uparrow} \\
				\bold{0}
			\end{array}\right),
		P\cvector^{-} = \left(\begin{array}{cc}
				\bold{0} \\
				\cvector^{\downarrow}
			\end{array}\right)
	\end{equation}
	where $\bold{dim}(\cvector^{\uparrow})+\bold{dim}(\cvector^{\downarrow}) = N$.
	Similarly, we set
	\begin{equation}
		\dvector^{+} = \frac{\dvector + \dprime}{2},
		\dvector^{-} = \frac{\dvector - \dprime}{2}.
	\end{equation}
	Then $\dvector^{-}_j$ is non-zero only if $\dvector^{+}_j$ is zero, and vice versa. Thus, by using another row swap matrix $Q\in \mathcal{A}(N)$,
	\begin{equation}
		\label{EQUATION-q-avector}
		Q\dvector^{+} = \left(\begin{array}{cc}
				\dvector^{\uparrow} \\
				\bold{0}
			\end{array}\right),
		Q\dvector^{-} = \left(\begin{array}{cc}
				\bold{0} \\
				\dvector^{\downarrow}
			\end{array}\right)
	\end{equation}
	where $\bold{dim}(\avector^{\uparrow})+\bold{dim}(\avector^{\downarrow}) = N$. From $\cprime = X\dprime$ and (\ref{EQUATION-relation}),
	\begin{equation}
		\cvector^{+} = X \dvector^{+}.
	\end{equation}
	By multiplying $P$ from left and using $Q^2=I$, we get
	\begin{equation}
		\label{EQUATION-p-cvector-plus}
		P\cvector^{+} = PXQ Q\dvector^{+}.
	\end{equation}
	Substituting the first equality in (\ref{EQUATION-p-cvector}) and that in  (\ref{EQUATION-q-avector}) into (\ref{EQUATION-p-cvector-plus}),
	\begin{equation}
		\label{EQUATION-up-matrix-form}
		\left(\begin{array}{cc}
				\cvector_{\uparrow} \\
				\bold{0}
			\end{array}\right) = \matrixsplit{X} \left(\begin{array}{cc}
				\dvector_{\uparrow} \\
				\bold{0}
			\end{array}\right)
	\end{equation}
	where we split $PXQ$ into submatrices so that the number of rows and columns of $X_{\uparrow\uparrow}$ is $\bold{dim}(\dvector^{\uparrow})$ and $\bold{ dim}(\cvector^{\uparrow})$ respectively. Writing the equality in (\ref{EQUATION-up-matrix-form}) explicitly, we get
	\begin{align}
		\cvector_{\uparrow} & = X_{\uparrow\uparrow} \dvector_{\uparrow},   \\
		\label{EQUATION-off-diagonal-condition-one}
		\bold{0}            & = X_{\downarrow\uparrow} \dvector_{\uparrow}.
	\end{align}
	Similarly, we obtain
	\begin{equation}
		\label{EQUATION-down-matrix-form}
		\left(\begin{array}{cc}
				\bold{0} \\
				\cvector_{\downarrow}
			\end{array}\right) = \matrixsplit{X} \left(\begin{array}{cc}
				\bold{0} \\
				\dvector_{\downarrow}
			\end{array}\right)
	\end{equation}
	and as a result,
	\begin{align}
		\label{EQUATION-off-diagonal-condition-two}
		\bold{0}              & = X_{\uparrow\downarrow} \dvector_{\downarrow},     \\
		\cvector_{\downarrow} & = X_{\downarrow\downarrow}   \dvector_{\downarrow}.
	\end{align}
	Since
	\begin{equation}
		\vectorsplit{\dvector} = Q\dvector,\ PXQ = \matrixsplit{X},
	\end{equation}
	the equations (\ref{EQUATION-off-diagonal-condition-one}) and (\ref{EQUATION-off-diagonal-condition-two}) indicate that $(\dvector, X)$ is pair-block-diagonal.
\end{proof}

From the theorem, it seems that when a dataset (hence $\dvector$) is given, we should find $X$ that $(\dvector, X)$ is {\it not} pair-block-diagonal; then by training $U(\boldsymbol{\theta})$ so that (\ref{EQUATION-casex-condition-1}) and (\ref{EQUATION-casex-condition-2}) with the $X$, the goal (\ref{EQUATION-goal}) is achieved. However, as far as our knowledge, for general $\dvector$, it is difficult to check if ($\dvector$, $X$) is pair-block-diagonal or not. On the other hand, if $\dvector_j \geq 0~(\forall j)$ or $\dvector_j \leq 0~(\forall j)$, we can show that $(\dvector, H^{\otimes n})$ is {\it not} pair-block-diagonal; although we already know the fact from the Theorem \ref{THEOREM-validity}, we can also prove it by directly showing that the condition for pair-block-diagonal is not satisfied when
$\dvector_j \geq 0~(\forall j)$ or $\dvector_j \leq 0~(\forall j)$ and $X=H^{\otimes n}$.
Therefore, instead of finding $X$ for general $\dvector$, we build our algorithm depending on the values of $\dvector$.

\begin{table*}[ht]
    \centering
    \caption{Stock prices for Exxon Mobil Corporation (XOM), Walmart (WMT), Procter \& Gamble (PG), Microsoft (MSFT), General Electronic (GE), AT\&T (T), Johnson \& Jonson (JNJ), and Chevron (CVX) between April 2008 and March 2009.}
    \label{TABLE-stock-data-2}
    \scalebox{1.1}{
    \begin{tabular}{|c|c|c|c|c|c|c|c|c|c|c|c|c|}
    \hline
			Symbol & Apr 08 & May 08 & Jun 08 & Jul 08 & Aug 08 & Sep 08 & Oct 08 & Nov 08 & Dec 08 & Jan 09 & Feb 09 & Mar 09 \\ \hline
			XOM    & 84.80  & 90.10  & 88.09  & 87.87  & 80.55  & 78.04  & 77.19  & 73.45  & 77.89  & 80.06  & 76.06  & 67.00  \\ \hline
			WMT    & 53.19  & 58.20  & 57.41  & 56.00  & 58.75  & 59.90  & 59.51  & 56.76  & 55.37  & 55.98  & 46.57  & 48.81  \\ \hline
			PG     & 70.41  & 67.03  & 65.92  & 60.55  & 65.73  & 70.35  & 69.34  & 64.72  & 63.73  & 61.69  & 54.00  & 47.32  \\ \hline
			MSFT   & 28.83  & 28.50  & 28.24  & 27.27  & 25.92  & 27.67  & 26.38  & 22.48  & 19.88  & 19.53  & 17.03  & 15.96  \\ \hline
        GE & 35.92 & 31.54 & 29.57 & 25.40 & 27.34 & 27.44 & 23.08 & 19.02 & 15.73 & 15.88 & 11.57 & 7.970 \\ \hline
        T & 38.70 & 39.29 & 39.67 & 33.41 & 31.01 & 32.53 & 28.15 & 26.87 & 28.00 & 28.74 & 24.97 & 22.80 \\ \hline
        JNJ & 65.13 & 67.13 & 66.55 & 63.75 & 68.50 & 71.09 & 69.07 & 61.49 & 57.66 & 60.13 & 57.25 & 49.03 \\ \hline
        CVX & 85.08 & 94.86 & 98.82 & 98.26 & 83.98 & 84.49 & 81.51 & 73.44 & 76.50 & 74.23 & 69.52 & 59.37 \\ \hline
    \end{tabular}
    }
\end{table*}

\section{Computation of the SVD entropy in the case of eight stocks}
\label{section:appendix-experiment}

Here we show the SVD entropy computation in a larger size setting. 
In addition to XOM, WMT, PG, and MSFT, we use the stock data of General 
Electronic (GE), AT\&T (T), Johnson \& Jonson (JNJ), and Chevron (CVX); 
they are top eight stocks included in the Dow Jones Industrial Average 
at the end of 2008. 
As in the case of TABLE \ref{TABLE-stock-data} in 
Section ~\ref{SECTION-demonstration}, we use the one-year monthly data 
from April 2008 to March 2009. 
Data was taken from Yahoo Finance (in every month, the opening price
is used). We show the stock price data in TABLE \ref{TABLE-stock-data-2}.

The goal is to compute the SVD entropy at each term, with the length $T=5$ months, which is the same as Section~\ref{SECTION-demonstration}.
The stock indices $j=1, 2, 3, 4, 5, 6, 7, 8$ correspond to XOM, WMT, PG, MSFT, GE, T, JNJ, and CVX, respectively.
Also, the time indices $t = 0, 1, 2, 3, 4$ identify the month in which the SVD entropy is computed; for instance, the SVD entropy on August 2008 is computed, using the data of April 2008 ($t=0$),
May 2008 ($t=1$), June 2008 ($t=2$), July 2008 ($t=3$), and August 2008 ($t=4$).
As a result, $s_{jt}$ has a total of $40 = 8~({\rm stocks}) \times 5~({\rm terms})$ components. Thus, 
from Eq.~\eqref{EQUATION-logarithm}, both $r_{jt}$ and $a_{jt}$ have $32 = 8 \times 4$ components,
where the indices run over $j=1, 2, 3, 4, 5, 6, 7, 8$ and $t = 1, 2, 3, 4$;
that is, ${\cal H}_{\rm stock}\otimes{\cal H}_{\rm time}={\bf C}^8\otimes{\bf C}^4$.
We use AAE algorithm in Case 2 for the data loading.
Hence, we need an additional ancilla qubit, meaning that the total number of qubit is 6.

Note that unlike the experiment in Section~\ref{SECTION-demonstration}, $n_s$ (=3) and $n_t$ (=2) are different, which requires a little modification for the cost function of the qSVD algorithm. 
Recall that the purpose of training PQCs $U_1(\bold{\xi})$ and $U_2(\bold{\xi^{\prime}})$ in qSVD is finding unitary operators that transform the Schmidt basis $\{|v_m\rangle_{\rm stock}\}_{m=1}^M$ and $\{|v_m^{\prime}\rangle_{\rm time}\}_{m=1}^M$ to the computational basis. We have freedom of the choice to which computational basis we transform the Schmidt basis, but we limit our goal of the training as 
$U_1(\bold{\xi})$ and $U_2(\bold{\xi^{\prime}})$ ideally operates as follows:
\begin{equation}
	\label{EQUATION-svd-state-modified}
	U_1(\xi) \otimes U_2(\xi^{\prime}) |\widetilde{Data}\rangle
	= \sum_{m=1}^{M} c_m(|\bar{m}\rangle |0\rangle) _{\stockdim} |\bar{m}\rangle_{\timedim}, 
\end{equation}
where $\{(|\bar{m}\rangle|0\rangle)_{\rm stock}\}_{m=1}^M$ is  a subset of the computational basis in $\mathcal{H}_{\rm stock}$ with the last qubit equal to zero and $\{|\bar{m}\rangle_{\rm time}\}_{m=1}^M$ is the subset of the computational basis in $\mathcal{H}_{\rm time}$. The corresponding cost function is given by
\begin{equation}
\label{EQUATION-qsvd-cost-modified}
	{\cal L}_{\rm SVD}(\xi, \xi')
	= \frac{1 - \sigma_z^{3}}{2} + \sum_{q=1}^{2} \frac{1 - \langle \sigma_z^{q} \sigma_z^{q+3}\rangle}{2},
\end{equation}
where the expectation $\langle \cdot \rangle$ is taken over
$U_1(\xi) \otimes U_2(\xi^{\prime}) |\widetilde{Data}\rangle$.
The operator $\sigma_z^q$ is the Pauli $Z$ operator that acts on the $q$-th qubit.
We can see that ${\cal L}_{\rm SVD}(\xi, \xi')=0$ holds, if and only if
$U_1(\xi) \otimes U_2(\xi^{\prime}) |\widetilde{Data}\rangle$ takes the form of the right hand side
of Eq.~\eqref{EQUATION-svd-state-modified}.
Therefore, by training $U_1(\xi)$ and $U_2(\xi^{\prime})$ so that ${\cal L}_{\rm SVD}(\xi, \xi')$ is minimized, we obtain the state that best approximates the Schmidt decomposed state.

The settings of AAE training is similar to the ones in Section~\ref{SECTION-demonstration}. 
As the PQC $U(\boldsymbol{\theta})$ for the data loading, we use the hardware-efficient ansatz with 6 qubits. The composition of each layer, the way of initialization, the kernel function, the optimizer, and the number of samples for computing $q_{\theta_r}^{\pm}$ and $q_{\theta_r}^{H\pm}$ are the same as the previous case in Section~\ref{SECTION-demonstration}. The learning rate is 0.1 for the first $100$ iterations and 0.01 for the other iterations. We chose the number of layers from 12 or 13. We performed 10 trials of training $U(\boldsymbol{\theta})$ for each number of layers (20 trials in total). Note that for each trial we saved the model at the $\{200, 250, 300, 350, 400\}$-th iterations and then the model which minimizes the cost $\mathcal{L}$ is chosen as the output of the trial. Among the outputs of each trial, the minimizer of $\mathcal{L}$ is adopted as the data loading circuit used for the computation of the SVD entropy in the next step. As a result, for computing the SVD entropy on August 2008, February 2009, and March 2009, 12-layers data loading circuits are used and for computing that on September 2008, October 2008, November 2008, December 2008, and January 2009, 13-layers data loading circuits are used.

\fig{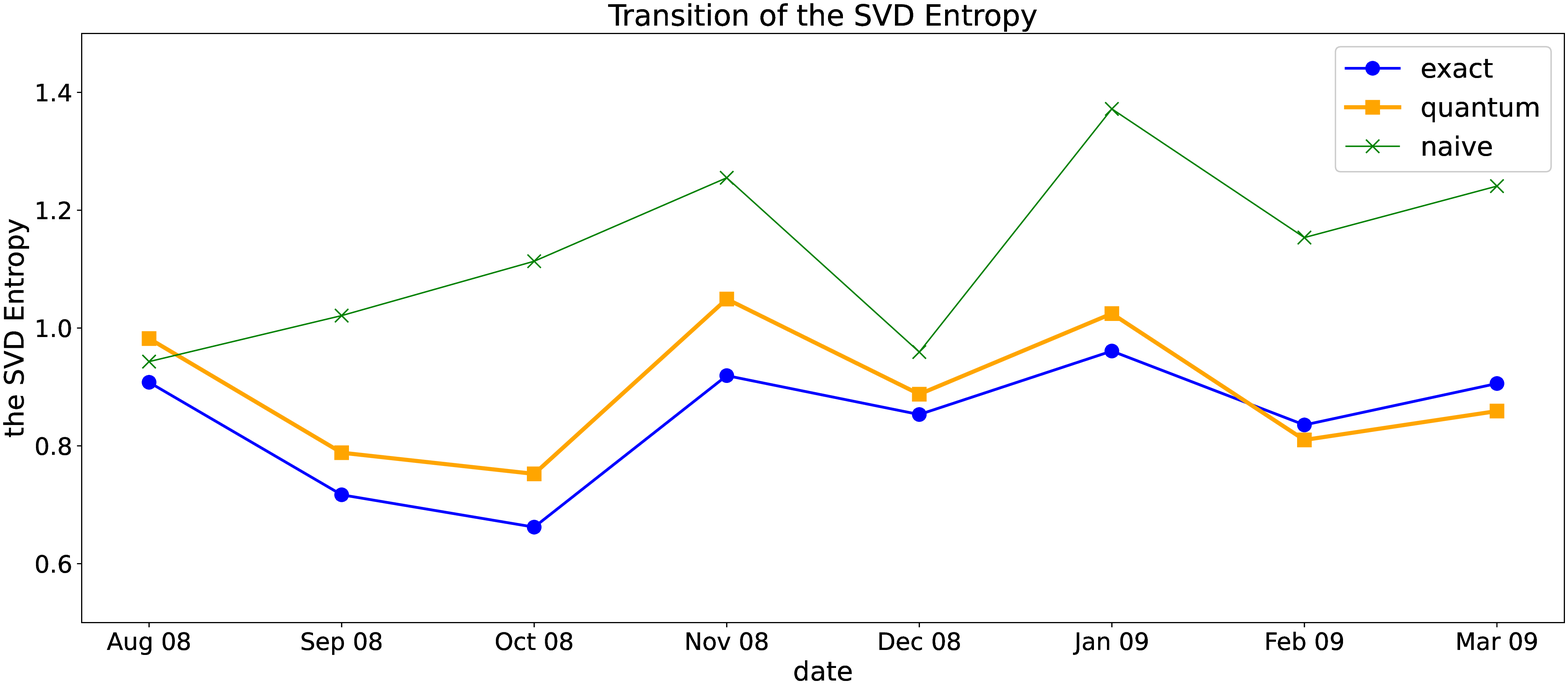}{
Change of the SVD entropy for each term, with different computing method when using stock data in TABLE \ref{TABLE-stock-data-2}.
The SVD entropy computed via AAE and qSVD algorithms, is shown by the orange
line with square dots. 
The exact value of SVD entropy, computed by diagonalizing the correlation matrix, is shown by the blue line with circle dots.
The SVD entropy computed with the naive data loading method is shown
by the green line with cross marks.}
{FIGURE-svd-entropy-2}
{width=500pt}

The qSVD is performed with the cost function \eqref{EQUATION-qsvd-cost-modified}.
The PQC $U_1(\bold{\xi})$, which acts on the stock state $|j\rangle$, is 
set to 3-qubits, 12-layers ansatz; also $U_2(\bold{\xi}^{\prime})$, which 
acts on the time state $|t\rangle$, is set to 2-qubits, 12-layers ansatz. 
The composition of the circuit, the way of initialization, the optimizer, 
the learning rate, and the number of iterations are the same as the 
previous qSVD experiment. 
The computation of the SVD entropy is also performed in the same way as 
in Section~\ref{SECTION-demonstration}. 

We show the SVD entropy in each term computed through AAE and qSVD algorithms by the orange line with square dots in Fig.~\ref{FIGURE-svd-entropy-2}.
As a reference, the exact value of SVD entropy, computed by diagonalizing the correlation matrix,
is shown by the blue line with circle dots.
Also, to see a distinguishing property of AAE, we study the naive data-loading method that trains the PQC $U_{\rm naive}(\boldsymbol{\theta})$ so that it learns only the absolute value of the data by minimizing the
cost function
$\mathcal{L}_{\rm MMD}(|\langle j|\langle  t|U_{\rm naive}(\boldsymbol{\theta})|0\rangle^{\otimes 5}|^2, a_{jt}^2)$.
The resulting value of SVD entropy computed with this naive method is shown by the green line with
cross marks.

Similar to the results in Section~\ref{SECTION-demonstration}, the SVD entropies computed with our AAE algorithm well approximate the exact values, while the naive method poorly works. Notably, despite the increase in the number of data, we see that the estimation errors in Fig.~\ref{FIGURE-svd-entropy-2} are about the same as those in Fig.~\ref{FIGURE-svd-entropy} (the maximum estimation error is about $10\%$ in both figures). Also, the number of layers for the data loading circuit in this Section  (=12 or 13) is smaller than twice that in Section~\ref{SECTION-demonstration} (=8) even though the number of data doubles and the number of qubits increases, which infers that the number of layers for AAE does not increase exponentially in conjunction with the increase of the number of qubits as expected. 
Still, as the number of qubits increases, the issues in the optimization discussed in Section~ \ref{SECTION-computational-complexity} may become severer; a larger size experiment is necessary to study how those issues affect our algorithm and how we can avoid them, which is beyond the scope of this paper and left for future work.
\end{document}